\documentclass[aps,pre,twocolumn,amsmath,amssymb,superscriptaddress,reprint,longbibliography]{revtex4-1}

\AtBeginDocument{
 \newwrite\bibnotes
 \def\bibnotesext{Notes.bib}
 \immediate\openout\bibnotes=\jobname\bibnotesext
 \immediate\write\bibnotes{@CONTROL{REVTEX41Control}}
 \immediate\write\bibnotes{@CONTROL{
apsrev41Control,author="08",editor="1",pages="1",title="0",year="1"}}
 \if@filesw
 \immediate\write\@auxout{\string\citation{apsrev41Control}}%
 \fi
}

\usepackage[utf8]{inputenc}
\usepackage{graphicx}
\usepackage{xcolor}
\usepackage[colorlinks=true,citecolor=blue]{hyperref}
\usepackage[caption=false]{subfig}
\usepackage{color}
\usepackage{cancel}
\usepackage{float}

\newcommand{\CS}{{\text{\tiny{CS}}}}
\newcommand{\MLE}{{\text{\tiny{MLE}}}}
\newcommand{\HR}{\hat{H}^{\text{\tiny{R}}}}
\newcommand{\tr}{\text{Tr}}

\begin{document}
%\begin{frontmatter}
\title{Entropy estimators for Markovian sequences: A comparative analysis}
\author{Juan De Gregorio}
\author{David S\'anchez}
\author{Ra\'ul Toral}
\affiliation{
Institute for Cross-Disciplinary Physics and Complex Systems IFISC (UIB-CSIC), Campus Universitat de les Illes Balears, E-07122 Palma de Mallorca, Spain
}
\date{\today}
\begin{abstract}
Entropy estimation is a fundamental problem in information theory that has applications in various fields, including physics, biology, and computer science. Estimating the entropy of discrete sequences can be challenging due to limited data and the lack of unbiased estimators. Most existing entropy estimators are designed for sequences of independent events and their performances vary depending on the system being studied and the available data size. In this work, we compare different entropy estimators and their performance when applied to Markovian sequences. Specifically, we analyze both binary Markovian sequences and Markovian systems in the undersampled regime. We calculate the bias, standard deviation, and mean squared error for some of the most widely employed estimators. We discuss the limitations of entropy estimation as a function of the transition probabilities of the Markov processes and the sample size. Overall, this paper provides a comprehensive comparison of entropy estimators and their performance in estimating entropy for systems with memory, which can be useful for researchers and practitioners in various fields.
\end{abstract}
%\end{frontmatter}
\maketitle

\section{\label{sec:level1} Introduction}

The entropy associated with a random variable is a measure of its uncertainty or diversity, taking large values for a highly unpredictable random variable (i.e., all outcomes equally probable) and low values for a highly predictable one (i.e., one or few outcomes much more probable than the others). As such, the concept has found multiple applications in a variety of fields including but not limited to nonlinear dynamics, statistical physics, information theory, biology, neuroscience, cryptography, and linguistics
\cite{Lewontin1972,stinson,neuroscience,biology,cover,letellier,victor,schindler,rosso,sherwin,e14081553,e19060275,e24040509}.

Due to its mathematical simplicity and clear interpretation, Shannon's definition is the most widely used measure of entropy~\cite{shannon}.
For a discrete random variable $X$ with $L$ distinct possible outcomes $x_1,\ldots,x_L$, the Shannon entropy reads
\begin{equation}
H[X] = -\sum_{i=1}^{L}p(x_i)\ln(p(x_i)),
\label{eq:entropy}
\end{equation}
where $p(x_i)$ denotes the probability that the random variable $X$ takes the value $x_i$.

It often occurs in practice that the probability distribution of the variable $X$ is unknown, either due to mathematical difficulties or to the lack of deep knowledge of the details of the underlying experiment described by the random variable $X$.
In those situations, it is not possible to compute the entropy using Equation~\eqref{eq:entropy} directly. In general, our information is restricted to 
a finite set of ordered data resulting from the observation of the outcomes obtained by repeating a large number of times, $N$, the experiment. Hence, the goal is to estimate $H$ from the ordered sequence $S = X_1,\ldots,X_N$, where each $X_j \in \lbrace x_i \rbrace_{i=1}^L$ with $j=1,\ldots,N$.

A numerical procedure that provides an approximation to the true value of $H$ based on the sequence $S$ is called an \textit{{entropy estimator}}. As the sequence $S$ is random, it is clear that an entropy estimator is itself a random variable, taking different values for different realizations of the sequence of $N$ outcomes. It would be highly desirable to have an unbiased entropy estimator, i.e., an estimator whose average value coincides with the true result $H$ for all values of the sequence length $N$. However, it can be proven that such an estimator does not exist~\cite{paninski} and that, apart from the unavoidable statistical errors due to the finite number $N$ of data of the sample (and which typically scale as $N^{-1/2}$), all estimators present systematic errors which are in general difficult to evaluate properly. Therefore, a large effort has been devoted to the development of entropy estimators that, although necessarily biased, provide a good value for $H$ with small statistical and systematic errors~\cite{contreras}.

The problem of finding a good estimator with small errors becomes more serious when the number of data $N$ is relatively small. Indeed, when the sizes of available data are much larger than the possible outcomes ($N \gg L$), it is not difficult to estimate $H$ accurately, and all of the most popular estimators are naturally satisfactory in this regime.
The task becomes much harder as the numbers $L$ and $N$ come closer to each other. It is particularly difficult in the 
undersampled regime ($N \lesssim L$)~\cite{undersample}, where some, or potentially many, possible outcomes may not be observed in the sequence. It is in this regime
where the difference in accuracy among the available estimators is more significant. 

We emphasize that the discussed difficulties already appear for independent identically distributed (i.i.d.) random variables. Precisely, the previous literature has largely dealt with entropy estimators proposed for sequences of i.i.d.~random variables~\cite{contreras,CS,Vu2007a,shr,arora}. However, it is not clear that real data arising from experimental observation can be described with i.i.d.~random variables due to the ubiquitous presence of data correlations. The minimal correlations in discrete sequences are of a Markovian nature. Then, how do the main entropy estimators behave for Markovian sequences?

The purpose of this work is to make a detailed comparison of some of the most widely used entropy estimators in systems whose future is conditionally independent of the past (Markovian). In Markovian sequences, correlations stem from the fundamental principle that the probability of a data value appearing at a specific time depends on the value observed in the preceding time step. Markov chains have been used to model systems in a large variety of fields such as statistical physics~\cite{gar04}, molecular biology~\cite{chu89}, weather forecast~\cite{wil99}, and linguistics~\cite{PhysRevLett.74.4559}, just to mention a few. 
Below, we analyze the strengths and weaknesses of estimators tested in a correlated series of numerically generated data. We compare the performances for the estimators that have shown to give good results for independent sequences~\cite{contreras}. For definiteness,
we below consider Markovian sequences of binary data.  {Furthermore,} 
  the calculation of relevant quantities in information theory, such as entropy rate and predictability gain~\cite{crutchfield2}, requires estimating the \textit{{block entropy}} of a sequence, obtained from the estimation of the entropy associated not to a single result, but to a block of consecutive results. As we will argue in the following sections, the construction of {overlapping} blocks induces correlations amongst them, even if the original sequence is not correlated. The calculation of the block entropy is also a tool that can be used to estimate the memory of a given sequence~\cite{juan}, which is of utmost importance when dealing with strongly correlated systems~\cite{yul06,ho12,sei12,sin14,mey21,WilsonKemsley2021}.

The rest of the paper is organized as follows. In Section~\ref{sec:estimators}, we make a brief overview of the {ten} entropy estimators being considered in this study{, nine of which are already known in the literature and an additional estimator built from results presented in ref.~\cite{econometrics7020017}, which is further developed in this work}. In Section~\ref{sec:results}, we present the results of our comparative analysis of these estimators in two Markovian cases: (A) binary sequences; and (B) in an undersampled regime. Section~\ref{sec:conclusions} contains the conclusions and an outlook. Finally, in Appendix~\ref{sec:appA} we provide a new interpretation in terms of geometric distributions of an estimator which is widely used as the starting point to construct others, and in Appendix~\ref{sec:appB} we prove the equivalence between a dynamics of block sequences and a Markovian random~\mbox{variable}.
%%%%%%%%%%%%%%%%%%%%%%%%%%%%%%%%%%%%%%%%%%
\section{\label{sec:estimators} Materials and Methods}

In the following, we will use the notation $\hat{a}$ to refer to a numerical estimator of the quantity $a$. The bias of $\hat{a}$ is defined as
\begin{equation}
B[\hat{a}] = \left\langle \hat{a} \right\rangle -a,
\label{eq:bias}
\end{equation}
{where $\left\langle \hat{a} \right\rangle$ represents the expected value of $\hat{a}$.}
The estimator $\hat{a}$ is said to be unbiased if $B[\hat{a}] = 0$. 
The dispersion of $\hat{a}$ is given by the standard deviation
\begin{equation}
\sigma[\hat{a}] = \sqrt{\langle \hat{a}^2 \rangle - \langle \hat{a} \rangle ^2}.
\label{eq:var}
\end{equation}
{Ideally,} 
 $\hat{a}$ should be as close to the true value $a$ as possible. Therefore, it is desirable that $\hat{a}$ has both low bias and low standard deviation. With this in mind, it is natural to consider the mean squared error of an estimator, given by
\begin{equation}
\text{MSE}[\hat{a}] = B[\hat{a}]^2 + \sigma[\hat{a}]^2,
\label{eq:mse}
\end{equation}
to assess its quality. Hence, when comparing estimators of the same variable, the one with the lowest mean squared error is preferable.

{Given an estimator $\hat{H}$ of the entropy, its $k$-th moment can be computed as
\begin{equation}\label{eq:hk}
\begin{split}
\langle \hat{H}^k\rangle=\sum_{S}P(S)\hat{H}(S)^k,
\end{split}
\end{equation}
where the sum runs over all possible sequences $S=X_1,\dots,X_N$ of length $N$ and $\hat{H}(S)$ is the value that the estimator takes on in this sequence. The probability $P(S)$ of observing the sequence $S$ depends on whether $S$ is correlated or not. For example, if $S$ is an independent sequence, $P(S)$ can be calculated as
\begin{equation}\label{eq:ps0}
 P(S) = \prod_{i=1}^{N} p(X_i).
\end{equation}
For correlated sequences, Equation~\eqref{eq:ps0} no longer holds. 
Consider a Markovian system, in which the probability of the next event only depends on the current state. In other words, the transition probabilities satisfy 
\begin{equation}\label{Markov}
\begin{split}
&P(X_s=x_j|X_{s-1}= x_{\ell},\ldots,X_1=x_k) = \\ &P(X_s=x_j|X_{s-1}= x_{\ell}),
\end{split}
\end{equation}
with $s$ the position in the series.
A homogeneous Markov chain is one in which the transition probabilities are independent of the time step $s$. Therefore, a homogeneous Markov chain is completely specified given the $L\times L$ matrix of transition probabilities $p(x_j| x_{\ell})=P(X_s=x_j|X_{s-1}= x_{\ell}),\,j,\ell=1,\dots,L$.
In this case, the probability of observing the sequence $S$ can be calculated as
\begin{equation}\label{eq:ps}
P(S) = p(X_1) \prod_{i=1}^{N-1} p(X_{i+1}|X_i). 
\end{equation}
where we have applied Equation~\eqref{Markov} successively.

The calculation of $P(S)$ can be generalized to an $m$-order Markov chain defined by the transition probabilities:
\begin{equation}\label{Markov-m}
\begin{split}
&P(X_s=x_j|X_{s-1}= x_{\ell},\ldots,X_1=x_k) = \\
&P(X_s=x_j|X_{s-1}= x_{\ell},\ldots,X_{s-m}=x_{u} ),
\end{split}
\end{equation}
that depend on the $m$ previous results of the random variable.

It is clear that the moments of the estimator $\hat{H}$, and consequently its performance given by its mean squared error, depend on the correlations of the system being analyzed. 

Most of the entropy estimators considered in this work only depend on the number of times each outcome occurs in the sequence. In this case, the calculation of the moments of the estimator can be simplified for independent and Markovian systems considering the corresponding multinomial distributions~\cite{wang1995markov}.}

Several entropy estimators were developed with the explicit assumption that the sequences being analyzed are uncorrelated~\cite{grassberger,bonachela}. The main assumption is that the probability of the number of times $n_i$ that the outcome $x_i$ occurs in a sequence of length $N$ follows a binomial distribution,
\begin{equation}
P(n_i) = \binom{N}{n_i}p(x_i)^{n_i}(1-p(x_i))^{N-n_i}.
\label{eq:binomial}
\end{equation} 
This approach is not valid when dealing with general Markovian sequences because \mbox{Equation~\eqref{eq:binomial}} no longer holds. {Instead, the Markovian binomial distribution \cite{markovbin} should be used, or more generally, the Markovian multinomial distribution \cite{wang1995markov}.}
Even for entropy estimators that were not developed directly using Equation~\eqref{eq:binomial}, their performance is usually only analyzed for independent sequences \cite{contreras}. Hence, the need to compare and evaluate the different estimators in Markov chains.

Even though there exists a plethora of entropy estimators in the literature~\cite{paninski,jackknife,wolpert1,vinck,zhang,archer,wolpert2,unseen,grassberger2,piga2023bayesian}, we here focus on nine of the most commonly employed estimators{, and we also propose a new estimator, constructed from known results \cite{econometrics7020017}.}

\subsection{Maximum Likelihood Estimator}
The maximum likelihood estimator (MLE) (also known as plug-in estimator) simply consists of replacing the exact probabilities in Equation~\eqref{eq:entropy} for the estimated frequencies,
\begin{equation}
\hat{p}(x_i) = \dfrac{\hat{n}_i}{N},
\label{eq:prob_mle}
\end{equation}
where $\hat{n}_i$ is the number of times that the outcome $x_i$ is {observed in the given sequence.} It is well known that Equation~\eqref{eq:prob_mle} is an unbiased estimator of $p(x_i)$, but the MLE estimator, given by
\begin{equation}
\hat{H}^\MLE = -\sum_{i=1}^{L} \hat{p}(x_i)\ln(\hat{p}(x_i)),
\label{eq:mle}
\end{equation}
is negatively biased~\cite{paninski}, i.e., $\langle \hat{H}^\MLE \rangle-H<0$.

\subsection{Miller--Madow Estimator} \label{sec:MM}
The idea behind the Miller--Madow estimator (MM)~\cite{MM} is to correct the bias of $\hat{H}^\MLE$ up to the first order in $1/N$, resulting in
\begin{equation}
\hat{H}^{\text{\tiny{MM}}} = \hat{H}^\MLE + \dfrac{N_0-1}{2N},
\label{eq:mm}
\end{equation}
where $N_0$ is the number of different elements present in the sequence. Corrections of higher order are not considered because they include the unknown probabilities $p(x_i)$~\citep{Sch_rmann_2004}.

\subsection{Nemenman--Shafee--Bialek Estimator}

A large family of entropy estimators are derived by estimating the probabilities using a Bayesian framework~\cite{minimax,1056331,SchG,bayes,wolpert1,wolpert2}.
The Nemenman--Shafee--Bialek estimator (NSB)~\mbox{\cite{nsb,nsb2,nsb3}} provides a novel Bayesian approach that, unlike traditional methods, does not rely on strong prior assumptions on the probability distribution. Instead, this method uses a mixture of Dirichlet priors, designed to produce an approximately uniform distribution of the expected entropy value. This ensures that the entropy estimate is not exceedingly biased by prior assumptions.

The {\tt Python} implementation developed in ref.~\cite{ndd} was used in this paper for the calculations of the NSB estimator.

\subsection{Chao--Shen Estimator}

The Chao--Shen estimator (CS)~\cite{CS} takes into account two corrections to Equation~\eqref{eq:mle} to reduce its bias: first, a Horvitz--Thompson adjustment~\cite{horvitz} to account for missing elements in a finite sequence; second, a correction to the estimated probabilities, {\linebreak}$\hat{p}^\CS(x_i)=\hat{C}^\CS\hat{p}(x_i)$, leading to
\begin{equation}
\hat{C}^\CS = 1-\dfrac{N_1}{N},
\end{equation}
where $N_1$ is the number of elements that appear only once in the sequence.

The Chao--Shen entropy estimator is then
\begin{equation}
\hat{H}^\CS = -\sum_{x_i \in S} \dfrac{\hat{p}^\CS(x_i)\ln(\hat{p}^\CS(x_i))}{1-(1-\hat{p}^\CS(x_i))^N}.
\label{eq:cs}
\end{equation}

\subsection{Grassberger Estimator}

Assuming that all $p(x_i) \ll 1$, the probability distribution of each $n_i$ can be approximated by a Poisson distribution. Following this idea, Grassberger (G) derived the estimator presented in ref.~\cite{grassberger} by first considering Rényi entropies of order $q$ \cite{renyi}:
\begin{equation}
H(q) = \dfrac{1}{q-1}\ln \sum_{i=1}^{L} p(x_i)^{q}.
\label{eq:renyi}
\end{equation}
Taking into account that the Shannon case can be recovered by taking the limit $q \rightarrow 1$, the author proposed a low bias estimator for the quantity $p^q$, for an arbitrary $q$. This approach led to the estimator given~by 
\begin{equation}
\hat{H}^{\text{\tiny{G}}} = \ln(N) - \dfrac{1}{N}\sum_{i=1}^{L} \hat{n}_i G_{\hat{n}_{i}},
\label{eq:gr}
\end{equation}
with $G_1 = -\gamma - \ln 2$, $G_2 = 2-\gamma - \ln 2 $, and the different values of $G_{\hat{n}_i}$ computed using the recurrence relation
\begin{align}
G_{2n+1} &= G_{2n} \\
G_{2n+2} &= G_{2n} + \dfrac{2}{2n+1},
\end{align}
where $\gamma=0.57721\dots$ is Euler's constant.

\subsection{Bonachela--Hinrichsen--Muñoz Estimator}

The idea behind the Bonachela--Hinrichsen--Muñoz estimator (BHM)~\cite{bonachela} is to make use of Equation~\eqref{eq:binomial} to find a balanced estimator of the entropy that, on average, minimizes the mean squared error. The resulting estimator is given by
\begin{equation}
\hat{H}^{\text{\tiny{BHM}}} = \dfrac{1}{N+2} \sum_{i=1}^L (\hat{n}_i+1) \sum_{j=\hat{n}_i+2}^{N+2} \dfrac{1}{j}.
\label{eq:bon}
\end{equation}

\subsection{Shrinkage Estimator}

The estimator proposed by Hausser and Strimmer~\cite{shr} (HS) is a shrinkage-type estimator~\cite{marvin}, in which the probabilities are estimated as an average of two models: 
\begin{equation}
\hat{p}^{\text{\tiny{HS}}}(x_i) = \alpha \dfrac{1}{L} + (1-\alpha)\hat{p}(x_i),
\end{equation}
where the weight $\alpha$ is chosen so that the resulting estimator $\hat{p}^{\text{\tiny{HS}}}$ has lower mean squared error than $\hat{p}$ and is calculated by~\cite{SchaferStrimmer}
\begin{equation}
\alpha = \text{min}\left(1,\dfrac{1-\sum_{i=1}^L (\hat{p}(x_i))^2}{(N-1)\sum_{i=1}^L (1/L-\hat{p}(x_i))^2}\right).
\end{equation}
Hence, the shrinkage estimator is
\begin{equation}
\hat{H}^{\text{\tiny{HS}}} = -\sum_{i=1}^L \hat{p}^{\text{\tiny{HS}}}(x_i)\ln(\hat{p}^{\text{\tiny{HS}}}(x_i)).
\label{eq:sh}
\end{equation}

\subsection{Chao--Wang--Jost Estimator}

The Chao--Wang--Jost estimator (CWJ)~\cite{chao} uses the series expansion of the logarithm function, as well as a correction to account for the missing elements in the sequence. This estimator is given by
\begin{align}
\hat{H}^{\text{\tiny{CWJ}}} &= \sum_{i=1}^L \dfrac{\hat{n}_i}{N}(\psi(N)-\psi(\hat{n}_i)) \\
&+\dfrac{N_1}{N}(1-A)^{1-N} \left(-\ln(A) - \sum_{j=1}^{N-1}\dfrac{1}{j}(1-A)^j\right),
\label{eq:chao}
\end{align}
where $\psi(z)$ is the digamma function and $A$ is given by
\begin{align}
\begin{split}
A = 
\begin{cases}
\dfrac{2N_2}{(N-1)N_1+2N_2}, \quad &\text{if }N_2 > 0, \\[3ex]
\dfrac{2}{(N-1)(N_1-1)+2}, \quad &\text{if }N_2=0, N_1 > 0, \\[3ex]
1, \quad &\text{if }N_1=N_2=0, 
\end{cases}
\end{split}
\end{align}
with $N_1$ and $N_2$ the number of elements that appear once and twice, respectively, in the sequence. %Please ensure original meaning retained.

In the supplementary material of ref.~\cite{chao}, it is proven that the first sum in \mbox{Equation~\eqref{eq:chao}} is the same as the leading terms of {the estimators developed in refs.~\cite{zhang,vinck}.}
In Appendix~\ref{sec:appA}, we show that each term in this sum is also equivalent to an estimator that takes into account the number of observations made prior to the occurrence of the element $x_i$.

\subsection{Correlation Coverage-Adjusted Estimator}

The correlation coverage-adjusted estimator (CC)~\cite{juan} uses the same ideas that support Equation~\eqref{eq:cs} but considers a different correction to the probabilities, $\hat{p}^{\text{\tiny{CC}}}(x_i)$ =$\hat{C}^{\text{\tiny{CC}}}\hat{p}(x_i)$, where now $\hat{C}^{\text{\tiny{CC}}}$ is calculated sequentially taking into account previously observed data,
\begin{equation}
\hat{C}^{\text{\tiny{CC}}} = 1-\sum_{j=1}^{N'}\dfrac{1}{N'+j}I(X_{N'+j} \notin (X_1,\ldots,X_{N'+j-1})),
\label{eq:Ccc}
\end{equation}
where $N' \equiv N/2$ and the function $I(Z)$ yields $1$ if the event $Z$ is true and $0$ otherwise. By construction, this probability estimator considers possible correlations in the sequence.

Then, the CC estimator is given by
\begin{equation}
\hat{H}^{\text{\tiny{CC}}} = -\sum_{x_i \in S} \dfrac{\hat{p}^{\text{\tiny{CC}}}(x_i)\ln(\hat{p}^{\text{\tiny{CC}}}(x_i))}{1-(1-\hat{p}^{\text{\tiny{CC}}}(x_i))^N}.
\label{eq:cc}
\end{equation}

{\subsection{Corrected Miller--Madow Estimator}
In ref.~\cite{econometrics7020017} it is shown that the bias of the MLE estimator can be approximated based on a Taylor expansion as
\begin{equation}
B[\hat{H}^{\MLE}] \approx -\dfrac{N_0-1}{2N}-\dfrac{1}{N} \sum_{l=1}^{\infty} K(l),
\label{eq:biasmle}
\end{equation}
where 
\begin{equation}
K(l) = \sum_{i=1}^L P(X_{s+l}=x_i|X_s=x_i) - 1.
\label{eq:K}
\end{equation}
Notice that the first term in Equation~\eqref{eq:biasmle} is simply the Miller--Madow correction shown in Section~\ref{sec:MM}, whereas the second term involves the unknown conditional probabilities with a lag $l$ that tends to infinity. These quantities can be hard to estimate directly from observations, especially if dealing with short sequences. However, the calculation of $K(l)$ can be simplified. Assuming that the sequence is independent, it can easily be seen that $K(l)=0$ for all $l$ and one recovers the Miller--Madow correction. Considering that the sequence is Markovian, then $K(l)$ can be written in a simpler way by first noticing that $P(X_{s+l}=x_j|X_s=x_i) = (\mathbb{T}^l)_{ij}$, where $\mathbb{T}$ is the $L\times L$ transition probability matrix given by $(\mathbb{T})_{ij} = p(x_j|x_i)$. Hence,
\begin{equation}
K(l) = \sum_{i=1}^L (\mathbb{T}^l)_{ii} -1 = \tr(\mathbb{T}^l)-1 = \sum_{i=1}^L \lambda_i^l - 1, 
\label{eq:K2}
\end{equation}
where $\tr(\mathbb{T}^l)$ is the trace of the matrix $\mathbb{T}^l$ and $\lambda_i$ are the eigenvalues of $\mathbb{T}$. The last equality of Equation~\eqref{eq:K2} is a well-known result in linear algebra. Given that $\mathbb{T}$ is a stochastic matrix, then all eigenvalues fulfil that $|\lambda|\leq 1$, and at least one eigenvalue is equal to $1$. We will assume that only $\lambda_1=1$ and we will discuss later on the case where more than one eigenvalue is equal to $1$.

We can write Equation~\eqref{eq:biasmle} as
\begin{equation}
B[\hat{H}^{\MLE}] \approx -\dfrac{N_0-1}{2N}-\dfrac{1}{N} \sum_{l=1}^{\infty} \sum_{i=2}^L \lambda_i^l.
\label{eq:biasmle2}
\end{equation}
Using the well-known result for the sum of the geometric series, then,
\begin{equation}
B[\hat{H}^{\MLE}] \approx -\dfrac{N_0-1}{2N}-\dfrac{1}{N} \sum_{i=2}^L \dfrac{\lambda_i}{1-\lambda_i}.
\label{eq:biasmle3}
\end{equation}
Notice that the convergence of the series of Equation~\eqref{eq:biasmle2} requires that none of the eigenvalues $\lambda_2,\dots,\lambda_L$ has an absolute value equal to $1$.

Given a finite sequence, we need to estimate the transition matrix $\mathbb{T}$ as
\begin{equation}
(\hat{\mathbb{T}})_{ij} = \hat{p}(x_j|x_i) = \dfrac{\hat{n}_{ij}}{\sum_{k=1}^{L}\hat{n}_{ik}}, 
\end{equation}
with $\hat{n}_{ik}$ the number of times the block $(x_i,x_k)$ is observed in the sequence. We can then calculate the eigenvalues $\hat{\lambda}_1,\ldots,\hat{\lambda}_L$ of the matrix $\hat{\mathbb{T}}$, which is also stochastic, and hence, one of its eigenvalues, $\hat{\lambda}_1$, is equal to $1$. Therefore, the proposed corrected Miller--Madow estimator (CMM) is
\begin{equation}
\hat{H}^{\text{\tiny{CMM}}} = \hat{H}^{\text{\tiny{MM}}} + \dfrac{1}{N}\sum_{i=2}^L \dfrac{\hat{\lambda}_i}{1-\hat{\lambda}_i}.
\label{eq:new}
\end{equation}
The correction to the MM estimator should only be used when the absolute value of all eigenvalues but $\hat{\lambda}_1$ of the stochastic matrix $\hat{\mathbb{T}}$ are not equal to $1$. Otherwise, it is recommended to avoid that correction and simply use $\hat{H}^{\text{\tiny{MM}}}$ as the estimator.}

%%%%%%%%%%%%%%%%%%%%%%%%%%%%%%%%%%%%%%%%%%
\section{\label{sec:results} Results}

We now proceed to compare the performance of the different estimators defined in the previous Section~\ref{sec:estimators}. Let us note first that, given a particular sequence, all entropy estimators, with the exception of the CC {and CMM estimators}, will yield exactly the same value if we permute arbitrarily all numbers in the sequence. The reason behind this difference is that although the CC estimator takes into account the order in which the different elements appear in the sequence{, and the CMM estimator considers the transition probabilities of the outcomes}, all other estimators are based solely on the knowledge of the number of times that each possible outcome appears, and this number is invariant under permutations. 

Certain estimators, such as CS or CC, can be calculated without any prior knowledge of the possible number of outcomes, $L$. This feature is particularly advantageous in fields like ecology, where the number of species in a given area may not be accurately known. Conversely, estimators like HS and NSB require an accurate estimate of $L$ for their~\mbox{computation.}

As mentioned before, when analyzing an estimator, there are two important statistics to consider: the bias and the standard deviation. Ideally, we would like an estimator with zero bias and low standard deviation. For the entropy, we have already argued that such an unbiased estimator does not exist. Hence, in this case, the ``best'' estimator (if it exists) would be the one that has the best balance between bias and standard deviation, i.e., the one with the lowest mean squared error given by Equation~\eqref{eq:mse}.

In this section, we will analyze and compare these three statistics---bias, standard deviation, and mean squared error---for the {ten} entropy estimators reviewed in Section~\ref{sec:estimators} in two main Markovian cases: (A) binary sequences; and (B) in an undersampled regime.

\subsection{Binary Sequences}

First, we consider homogeneous Markovian binary ($L=2$) random variables, with possible outcomes $x_i = 0,1$. One advantage of discussing this system is that it is uniquely defined by a pair of independent transition probabilities, $p(0|0)$ and $p(1|1)$, where $p(x_i|x_j) \equiv P(X_{s+1}=x_i|X_s=x_j)$. Then, $p(1|0)=1-p(0|0)$ and $p(0|1)=1-p(1|1)$. To shorten the notation, we hereafter write $p_{00}$ for $p(0|0)$ and $p_{11}$ for $p(1|1)$.

It is possible to compute the Shannon entropy of this random variable using the general definition given by Equation~\eqref{eq:entropy}.
\begin{equation}\label{eq:H_binary}
H=-p(0)\ln p(0)-p(1)\ln p(1)
\end{equation}
with the stationary values \cite{cover}:{
\begin{equation}\label{pst}
\begin{split}
p(0) &= \dfrac{1-p_{11}}{2-p_{00}-p_{11}},\\
p(1)&=1-p(0).
\end{split}
\end{equation}}

{The average value and standard deviation of the different entropy estimators were computed using Equation~\eqref{eq:hk} for $k=1,2$ by generating all $2^N$ possible sequences $S$ and computing the probability of each one using Equation~\eqref{eq:ps}, where $p(X_1)$ are the stationary values given by Equation~\eqref{pst}.
We have followed this approach to compute the estimator bias $B=\langle \hat{H}\rangle-H$ and its standard deviation $\sigma=\sqrt{\langle \hat{H}^2\rangle-\langle \hat{H}\rangle^2}$. As an example, we plot the absolute value of the bias for sequences of length $N=4$ in the colour map of Figure~\ref{fig:bias}, for the {ten} entropy estimators presented in Section~\ref{sec:estimators}, as a function of the transition probabilities $p_{00}$ and $p_{11}$.

\begin{figure*}
 \center{\subfloat{\includegraphics[width=0.3\textwidth]{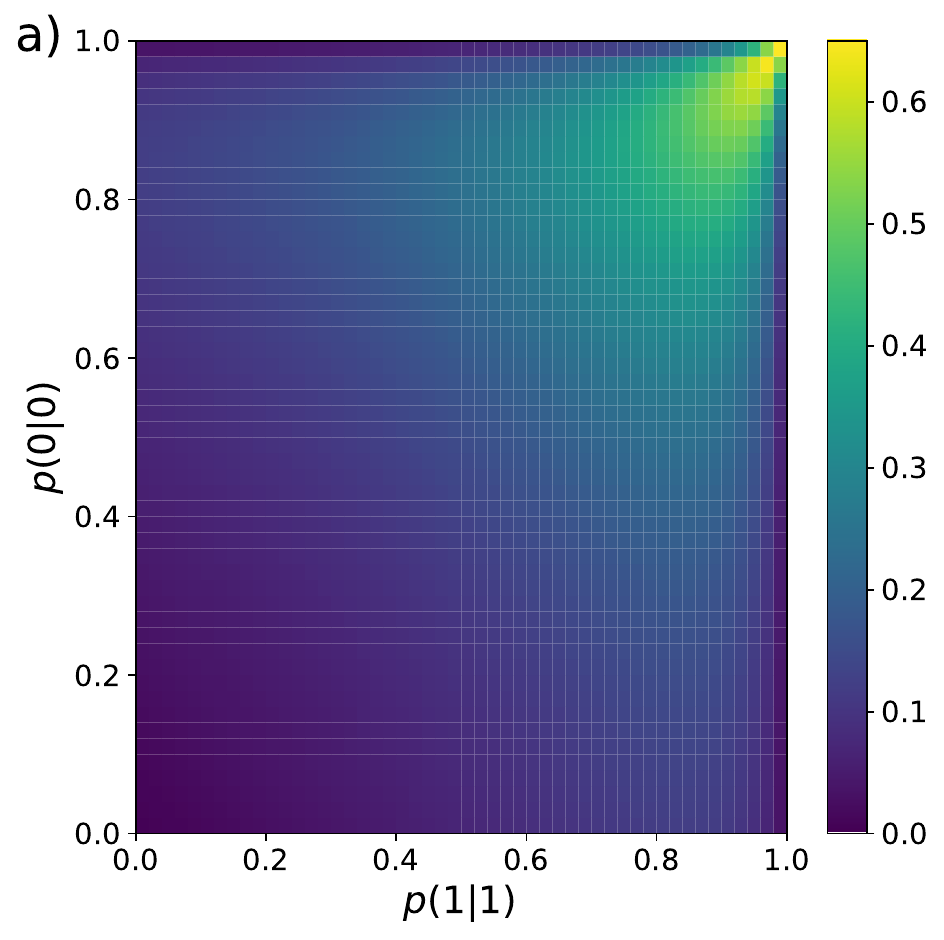}}} \\
	\subfloat{\includegraphics[width=0.3\textwidth]{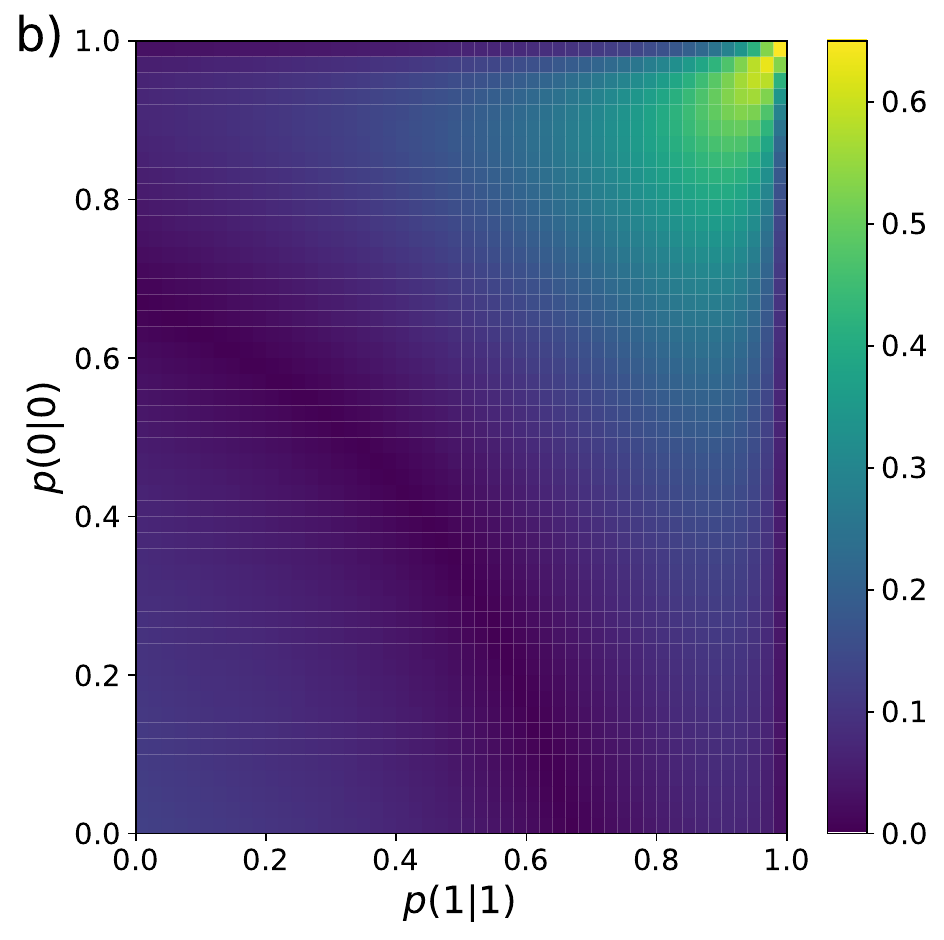}}
	\subfloat{\includegraphics[width=0.3\textwidth]{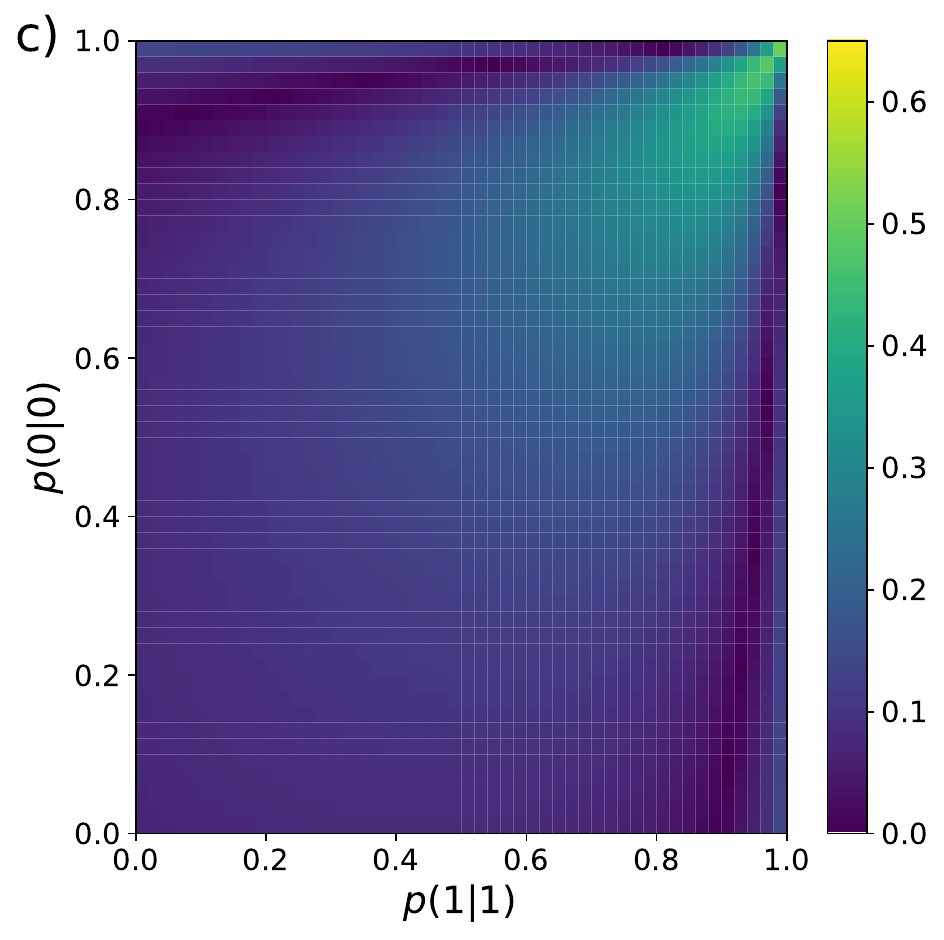}} 
	\subfloat{\includegraphics[width=0.3\textwidth]{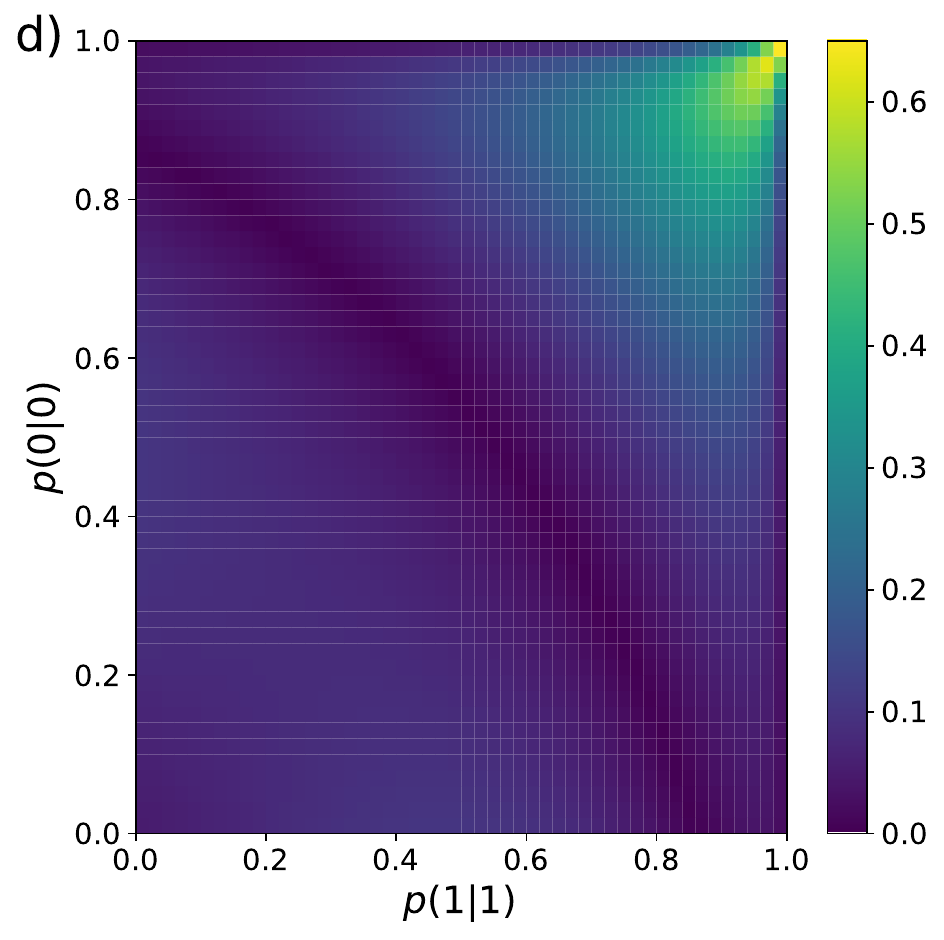}} \\
 %\newline
	\subfloat{\includegraphics[width=0.3\textwidth]{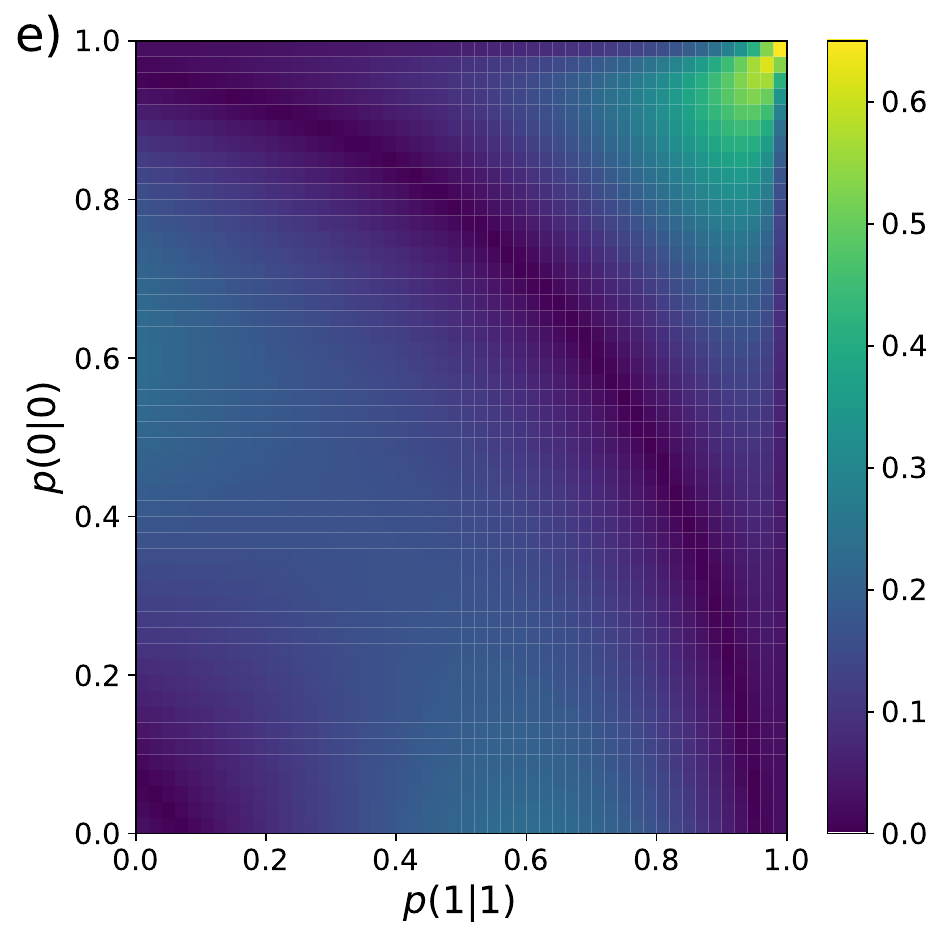}}
	\subfloat{\includegraphics[width=0.3\textwidth]{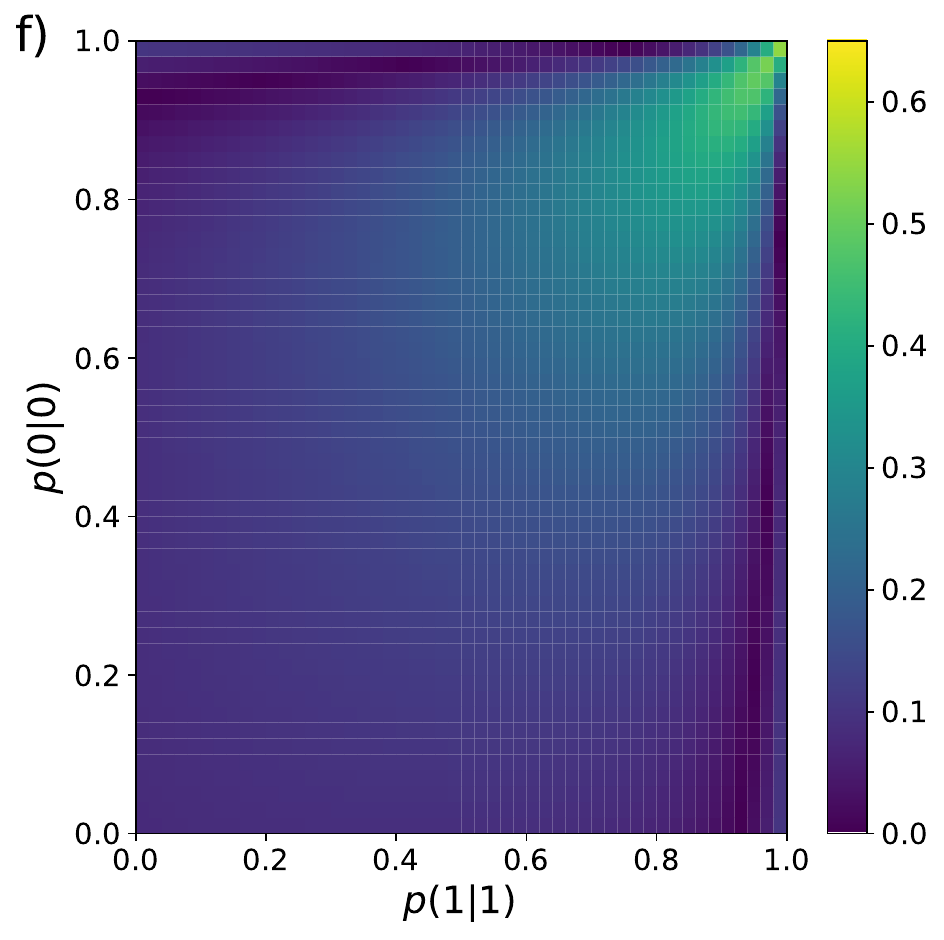}} 
	\subfloat{\includegraphics[width=0.3\textwidth]{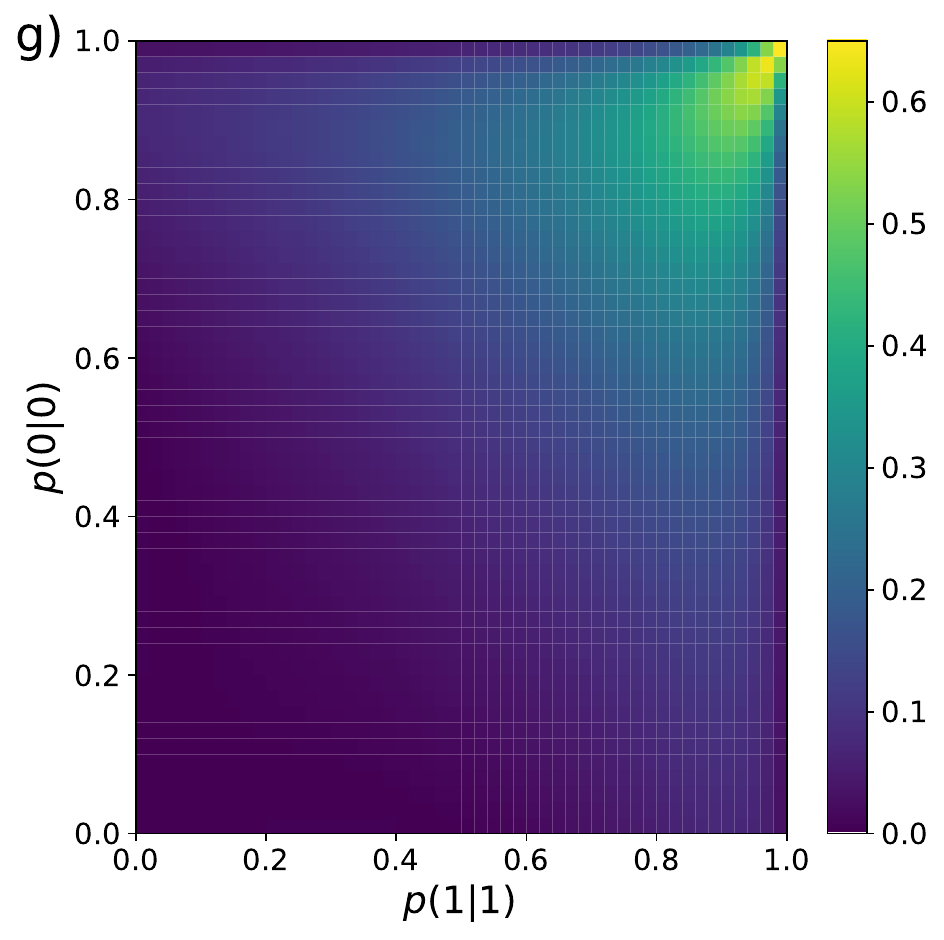}} \\
	\subfloat{\includegraphics[width=0.3\textwidth]{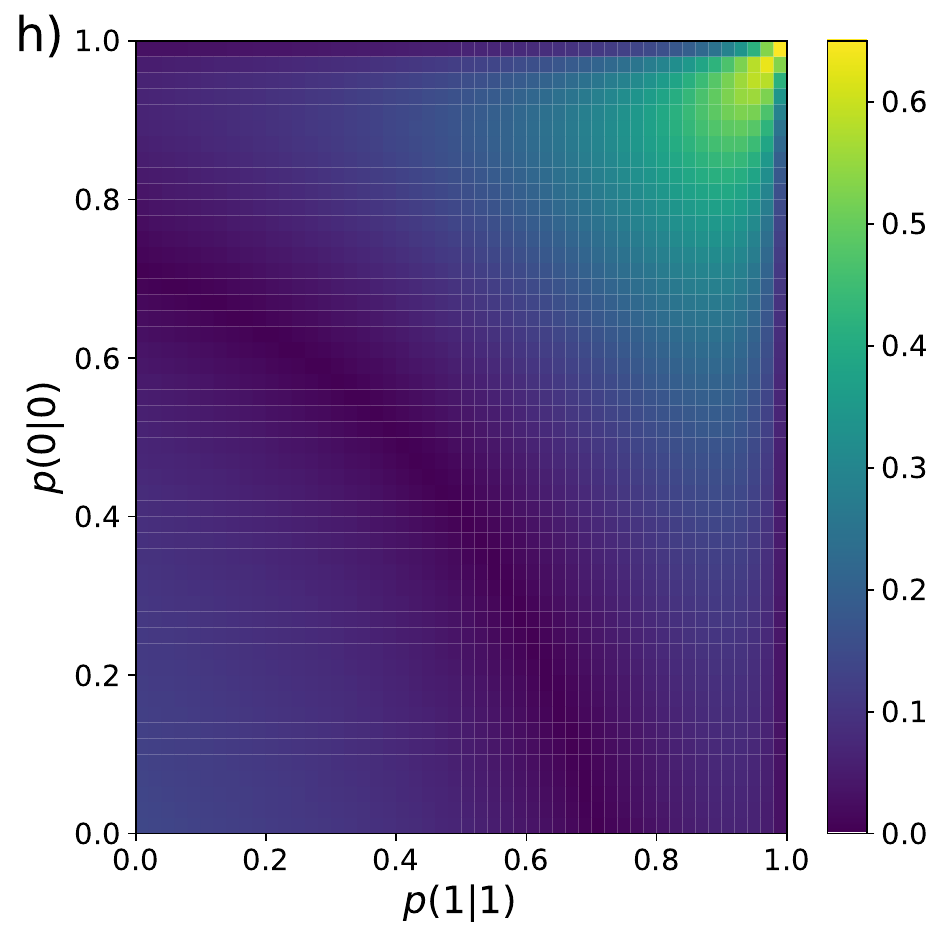}}
	\subfloat{\includegraphics[width=0.3\textwidth]{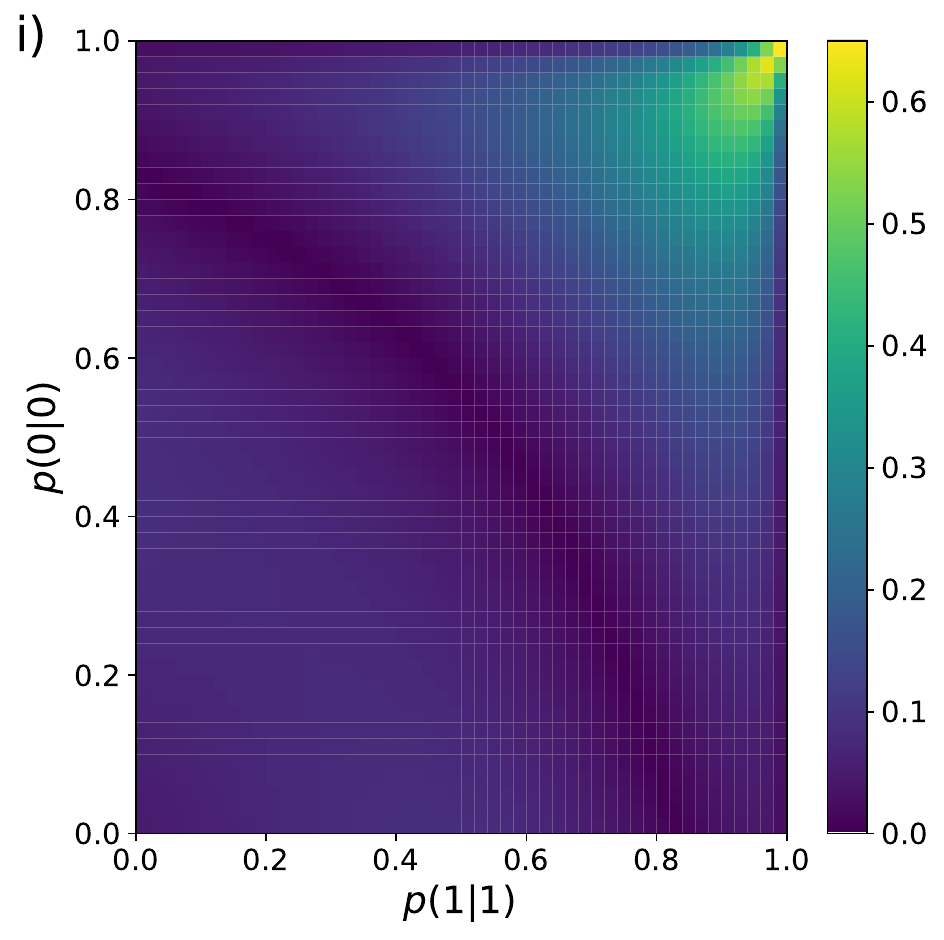}}
 \subfloat{\includegraphics[width=0.3\textwidth]{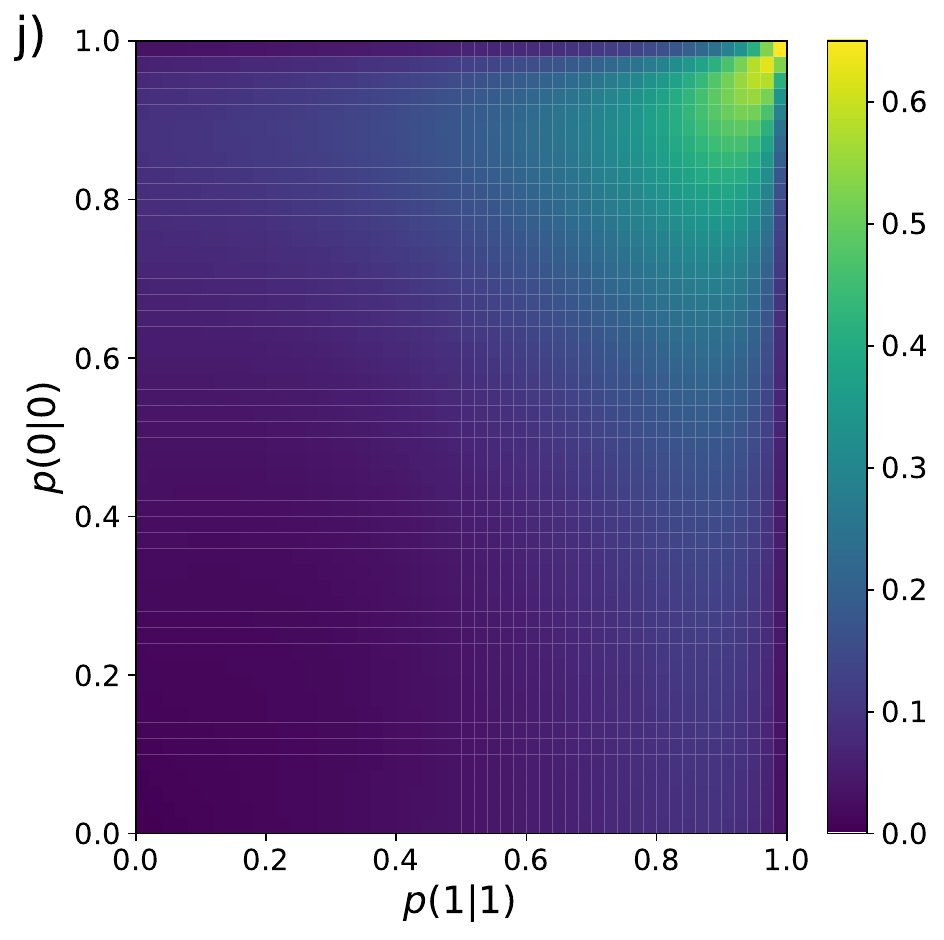}}
\caption {Colour maps representing the bias of the nine entropy estimators reviewed in Section~\ref{sec:estimators} for Markovian binary sequences of length $N=4$. The values of the transition probabilities $p(0|0)$ and $p(1|1)$ vary from $0.01$ to $0.99$ with step $\Delta p = 0.02$. (\textbf{a}) MLE [Eq.~\eqref{eq:mle}], (\textbf{b}) Miller--Madow \cite{MM}, (\textbf{c}) Nemenman et~al.~\cite{nsb}, (\textbf{d})  Chao--Shen \cite{CS}, (\textbf{e}) Grassberger \cite{grassberger}, (\textbf{f}) Bonachela et~al.~\cite{bonachela}, (\textbf{g}) Shrinkage \cite{shr}, (\textbf{h}) Chao et~al. \cite{chao}, (\textbf{i}) correlation coverage-adjusted \cite{juan}, (\textbf{j}) corrected Miller--Madow [Eq.~\eqref{eq:new}].}
\label{fig:bias}
\end{figure*}

In Figure~\ref{fig:bias}, we can see that, for all ten estimators, the bias is larger in the region around the values $p_{00} \simeq p_{11} \simeq 1$. The reason is that, in this region, the stationary probabilities of $0$ and $1$ are very similar, but given these particular values of the transition probabilities, a short sequence will most likely feature only one of these values, which makes it very hard to correctly estimate the entropy in those cases. Apart for this common characteristic, the performance of the estimators when considering only the bias is quite diverse, all of them having different regions where the bias is lowest (darker areas in the~panels). 

In order to quantitatively compare the performance of the different estimators, we have aggregated all values in the $(p_{00},p_{11})$ plane. We define the aggregated bias of an~\mbox{estimator,}
\begin{equation}\label{eq:B}
\overline{B} = (\Delta p)^2\sum_{p_{00},p_{11}}|B(p_{00},p_{11})| ,
\end{equation}
where the sum runs over all values of the transition probabilities used to produce Figure~\ref{fig:bias}, $\Delta p=0.02$ is the step value used for the grid of the figure, and $B(p_{00},p_{11})$ is the bias for the particular values of the transition probabilities. 
The aggregated bias given by Equation~\eqref{eq:B} depends only on the sequence length $N$.

We conduct the previous analysis for different values of $N$. The resulting plot of the aggregate bias $\overline{B}$ of the entropy estimator as a function of the sequence length is shown in Figure~\ref{fig:bias_total}. In this figure, we can see that {the CC estimator gives the best performance for small values of $N$, except for $N=2$, where the CWJ estimator has the lowest aggregated bias. However, from $N=7$ it is the CMM estimator which outperforms the rest. The poor performance of this estimator for low values of $N$ is due to the fact that this estimator, in contrast to the others, requires estimating the transition probabilities, as well as the stationary probabilities, and therefore more data are needed.} As expected, all the estimators yield an aggregated bias that vanishes as $N$ increases.

In the colour map of Figure~\ref{fig:sigma}, we perform a similar analysis for the standard deviation $\sigma$. In the figure, we find that all ten estimators show a similar structure in the sense that the regions of lowest and highest $\sigma$ are alike. The smallest deviation is mostly located near the left bottom corner of the colour maps and the largest deviation occurs around the regions $(0.65 \lesssim p_{00}\lesssim 0.9,\,0 \lesssim p_{11}\lesssim 1)$ and $(0 \lesssim p_{00}\lesssim 1,\,0.65 \lesssim p_{11}\lesssim 0.9)$ (green areas in the figures). Of course, the values of $\sigma$ inside these regions vary for each estimator but they all share this similar feature. In this case, by just looking at the colour maps, it is easy to see that BHM (panel f) and NSB (panel c) estimators are the ones with the lowest standard~deviation. 

\begin{figure}[!h]
\includegraphics[width=1\columnwidth]{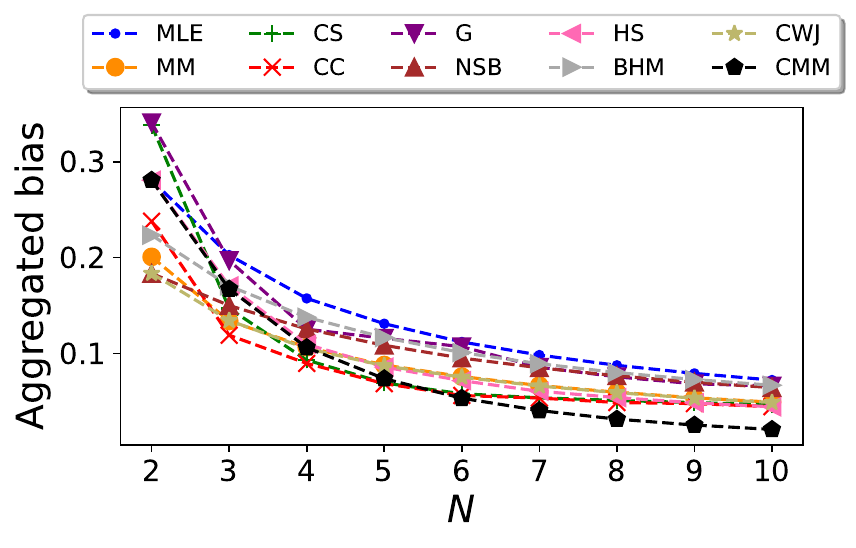}
\caption{Aggregated bias of the entropy estimators for Markovian binary sequences as a function of the sequence size $N$.}
\label{fig:bias_total}
\end{figure}
\unskip

\begin{figure*}
 \center{
 \subfloat{\includegraphics[width=0.3\textwidth]{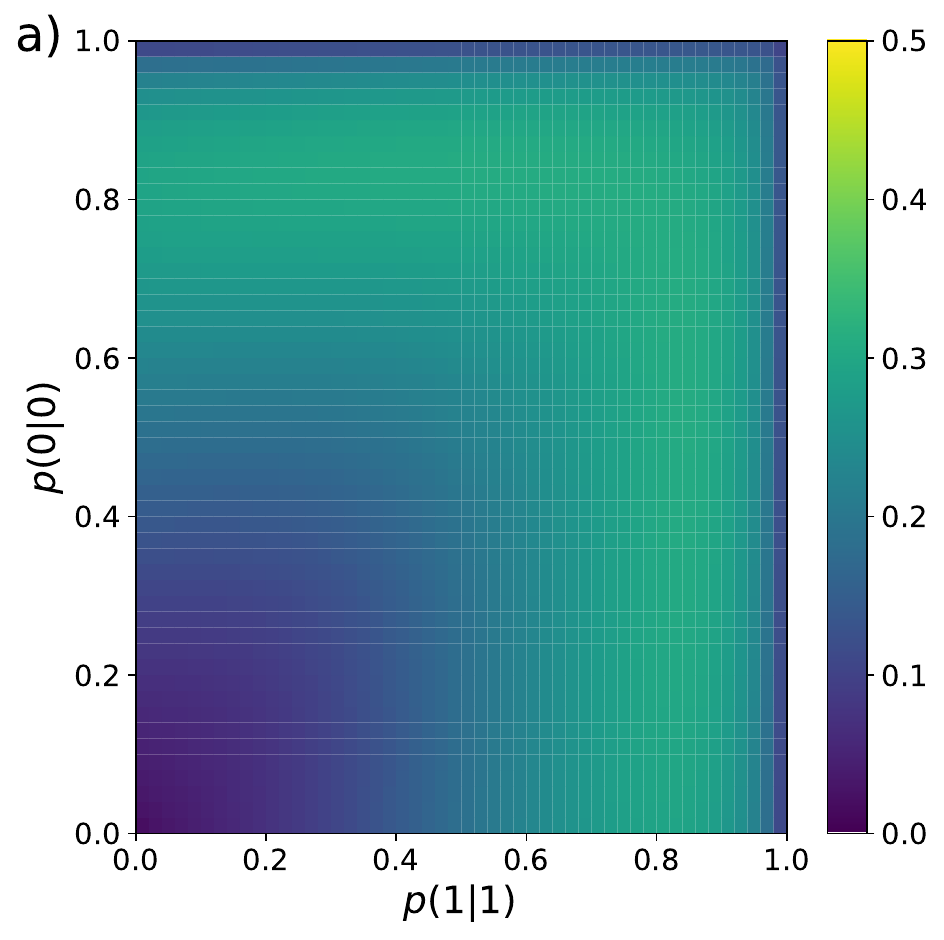}}} \\
%\newline
\subfloat{\includegraphics[width=0.3\textwidth]{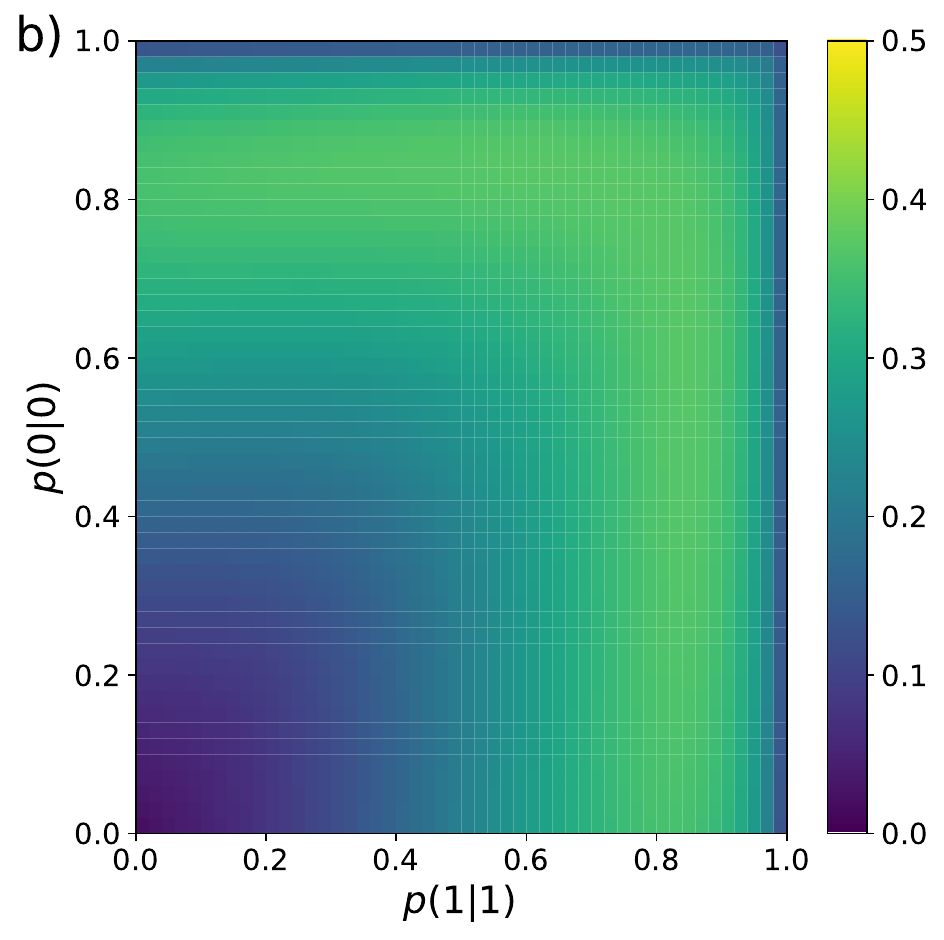}}
\subfloat{\includegraphics[width=0.3\textwidth]{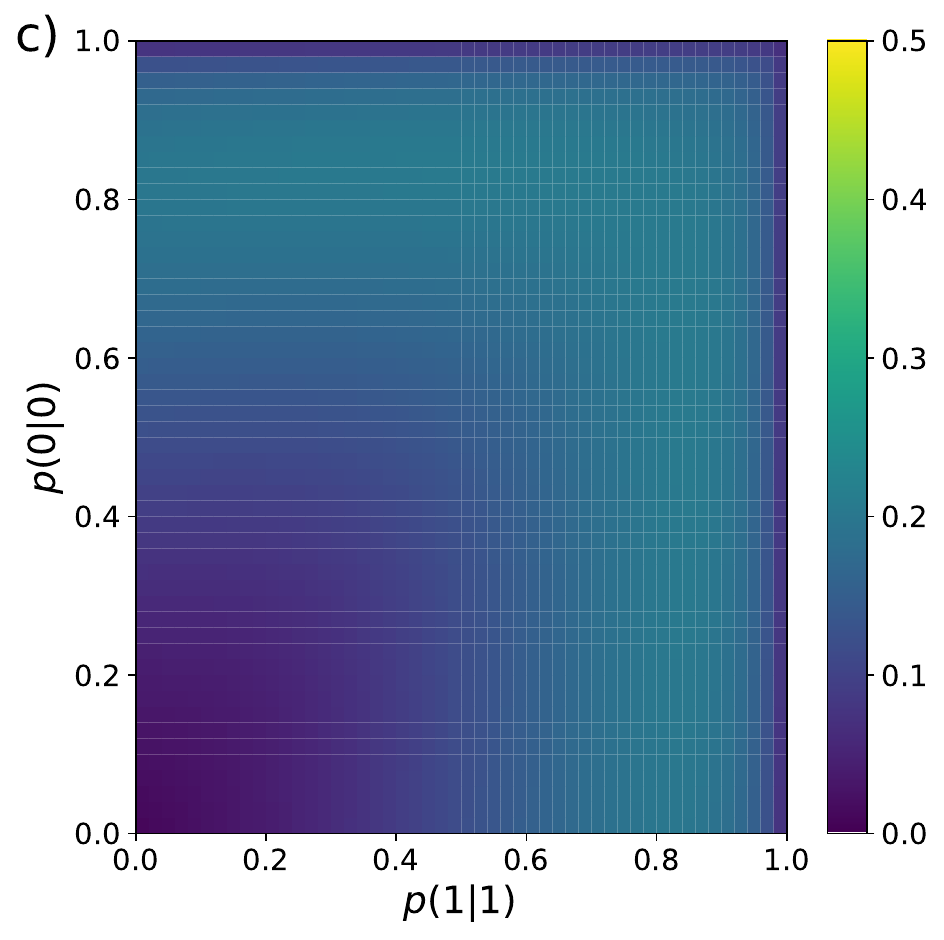}} 
\subfloat{\includegraphics[width=0.3\textwidth]{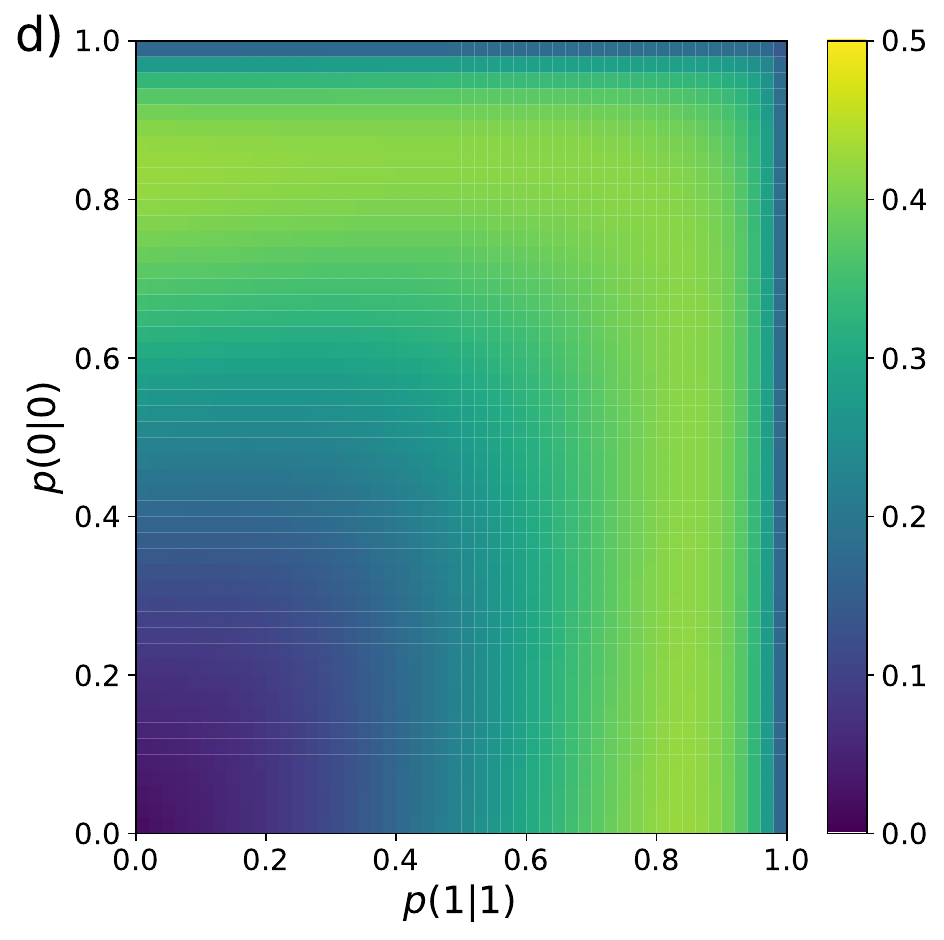}} \\
\subfloat{\includegraphics[width=0.3\textwidth]{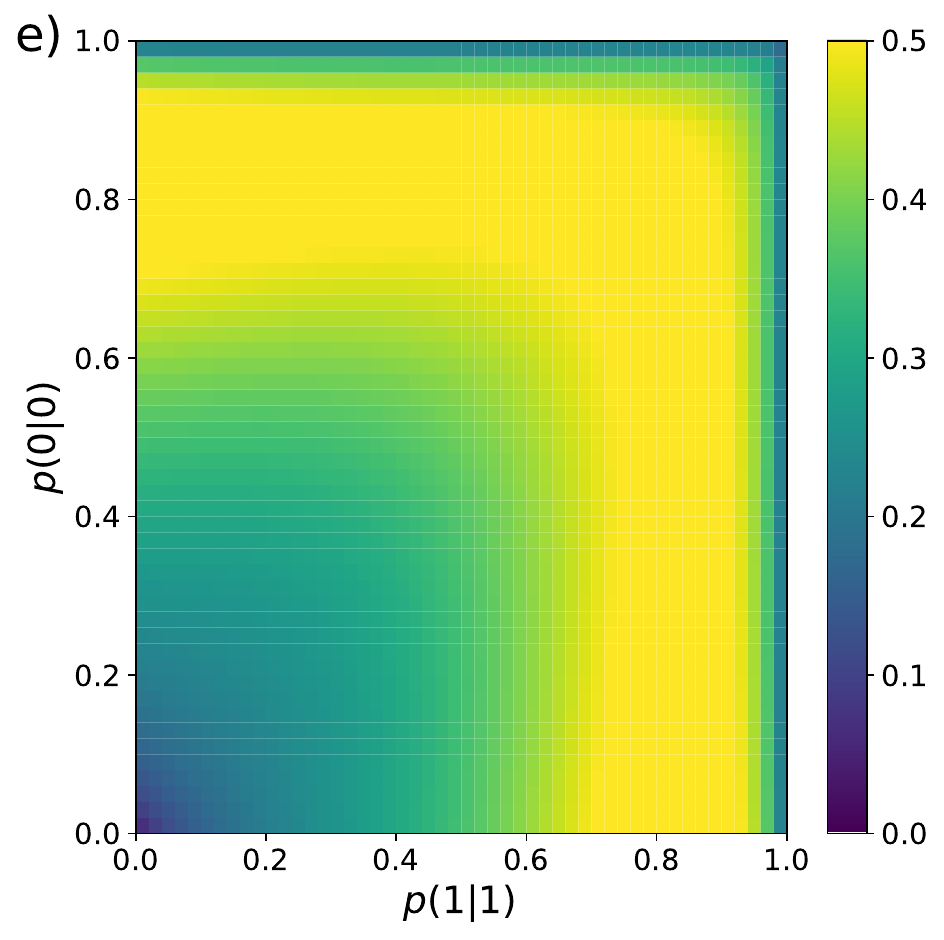}}
\subfloat{\includegraphics[width=0.3\textwidth]{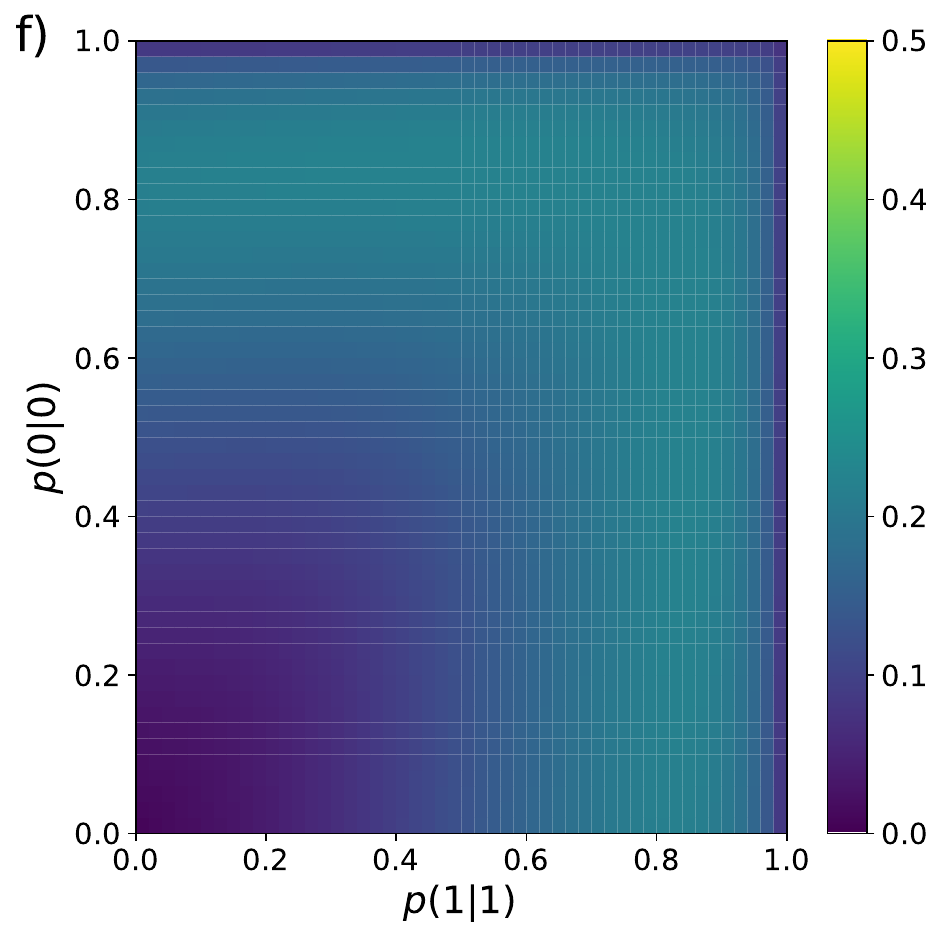}} 
\subfloat{\includegraphics[width=0.3\textwidth]{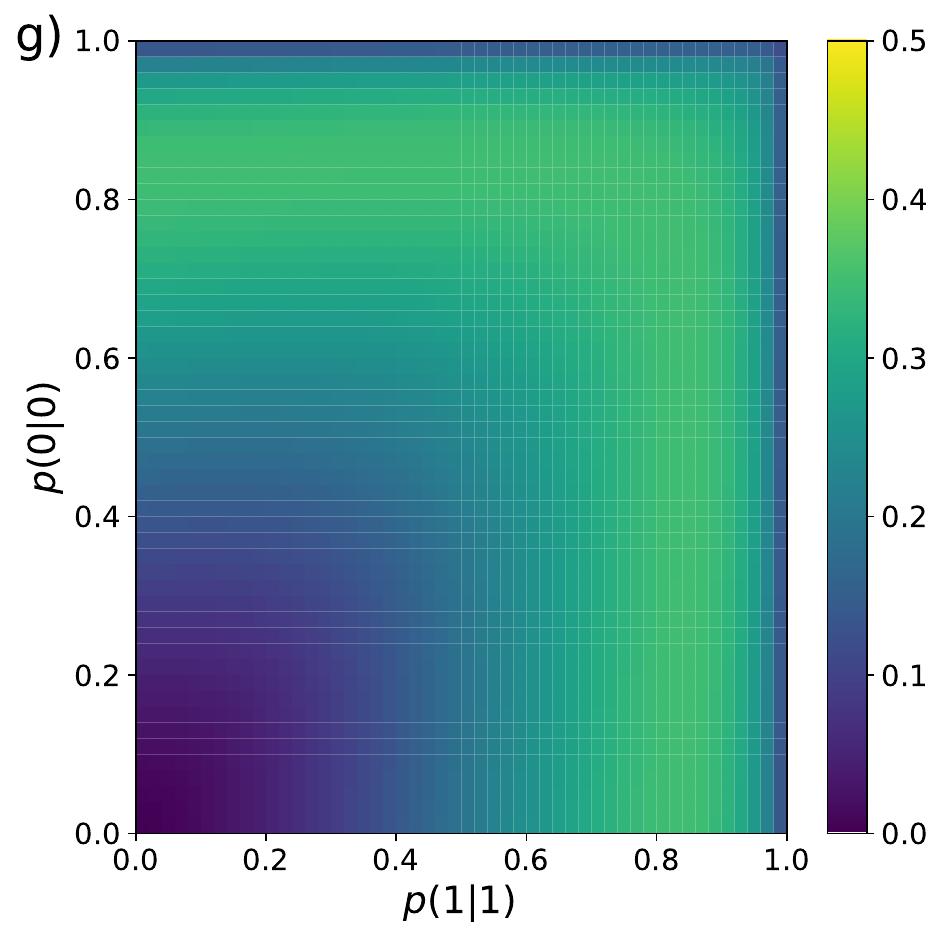}} \\
\subfloat{\includegraphics[width=0.3\textwidth]{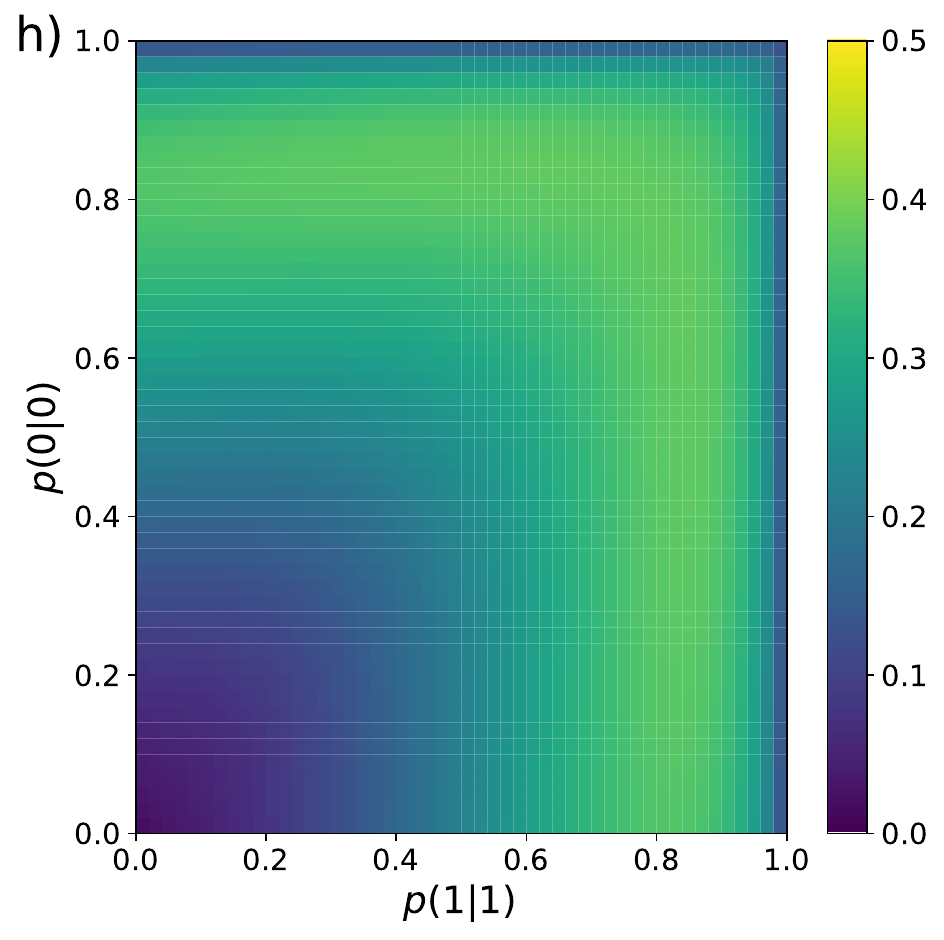}}
\subfloat{\includegraphics[width=0.3\textwidth]{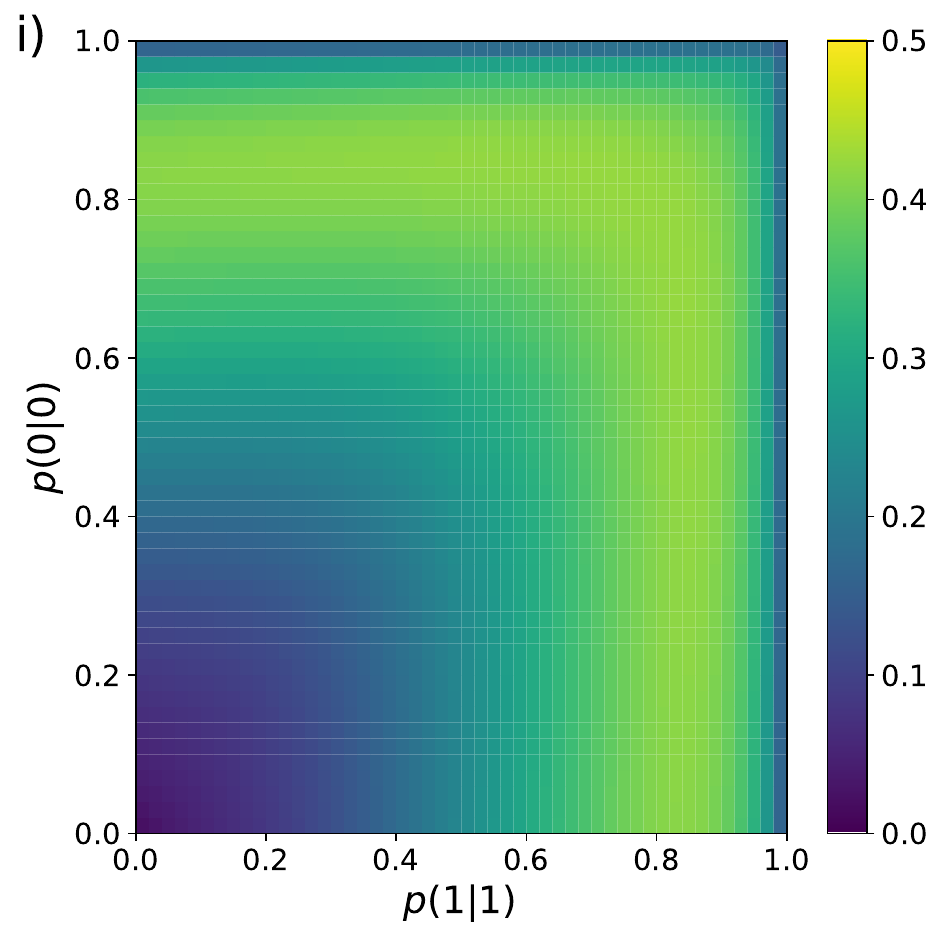}}
 \subfloat{\includegraphics[width=0.3\textwidth]{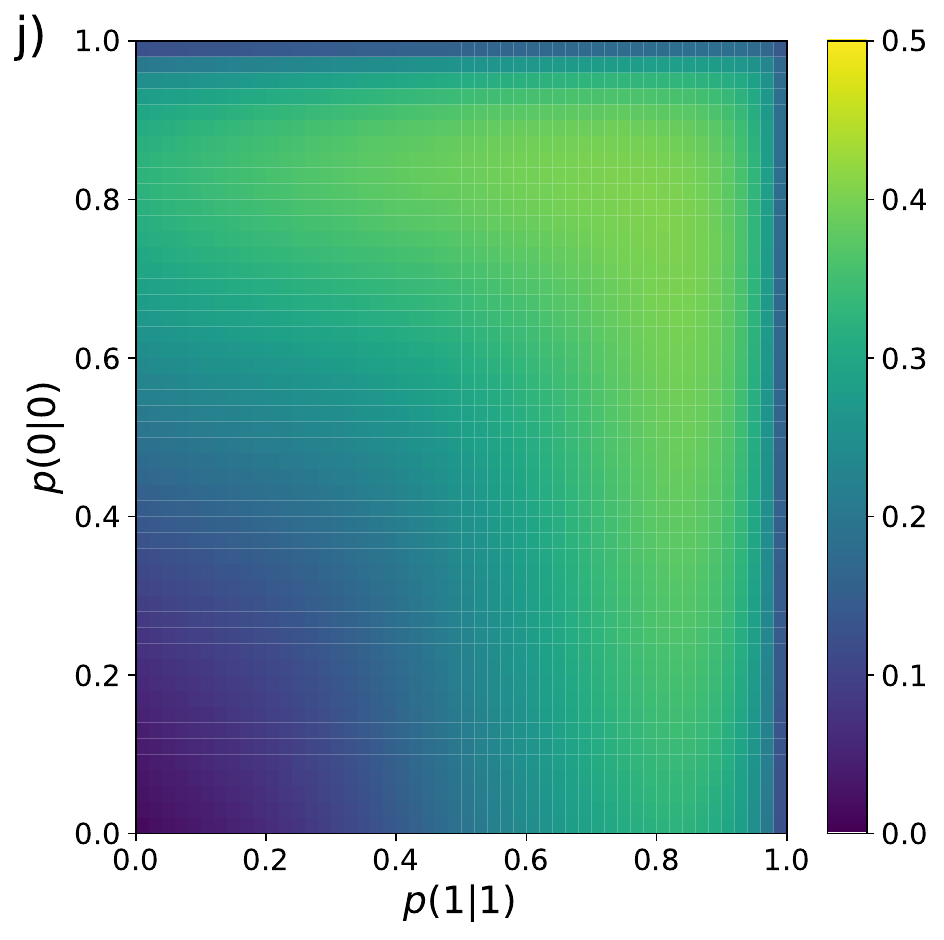}}
\caption{Colour maps representing the standard deviation of the nine entropy estimators reviewed in Section~\ref{sec:estimators} for Markovian binary sequences of length $N=4$. The values of the transition probabilities $p(0|0)$ and $p(1|1)$ vary from $0.01$ to $0.99$ with step $\Delta p = 0.02$. (\textbf{a}) MLE [Eq.~\eqref{eq:mle}], (\textbf{b}) Miller--Madow \cite{MM}, (\textbf{c}) Nemenman et~al.~\cite{nsb}, (\textbf{d})  Chao--Shen \cite{CS}, (\textbf{e}) Grassberger \cite{grassberger}, (\textbf{f}) Bonachela et~al.~\cite{bonachela}, (\textbf{g}) Shrinkage \cite{shr}, (\textbf{h}) Chao et~al. \cite{chao}, (\textbf{i}) correlation coverage-adjusted \cite{juan}, (\textbf{j}) corrected Miller--Madow [Eq.~\eqref{eq:new}].}
\label{fig:sigma}
\end{figure*}

The aggregated standard deviation $\overline{\sigma}$, defined in a similar way to the aggregated bias,
\begin{equation}\label{eq:sigma}
\overline{\sigma} = (\Delta p)^2\sum_{p_{00},p_{11}}\sigma(p_{00},p_{11}),
\end{equation}
is plotted in Figure~\ref{fig:sigma_total} as a function of the sequence size $N$. In agreement with the previous visual test, the BHM and NSB estimators clearly outperform the rest, even though their advantage is less significant as $N$ increases.

Finally, for every particular $N$, we compute the mean squared error of the entropy estimators, Equation~\eqref{eq:mse}, as a function of $p_{00}$ and $p_{11}$. Its aggregated value 
\begin{equation}\label{eq:MSE}
\overline{\text{MSE}} = (\Delta p)^2\sum_{p_{00},p_{11}}\text{MSE}(p_{00},p_{11}),
\end{equation}
is plotted as a function of $N$ in Figure~\ref{fig:mse_total}. Even though the {CC and CMM estimators outperform the others when considering only the bias, their large dispersion dominates the mean squared error.} Overall, it can be seen that the BHM and NSB estimators surpass the rest when both the bias and standard deviation are considered although, again, their advantage becomes less significant as $N$ increases.

\begin{figure}[H]
\includegraphics[width=1\columnwidth]{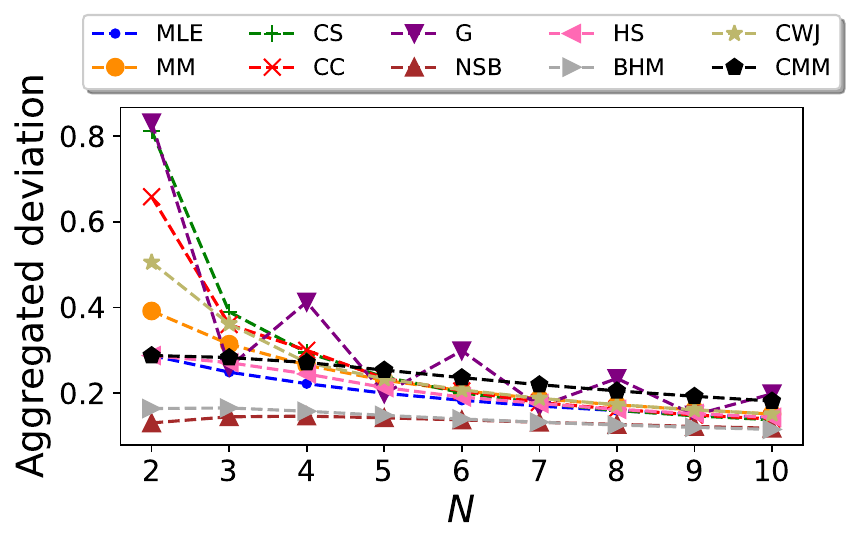}
\caption{Aggregated standard deviation of the entropy estimators for Markovian binary sequences as a function of the sequence size $N$.}
\label{fig:sigma_total}
\end{figure}
\unskip

\begin{figure}[H]
\includegraphics[width=1\columnwidth]{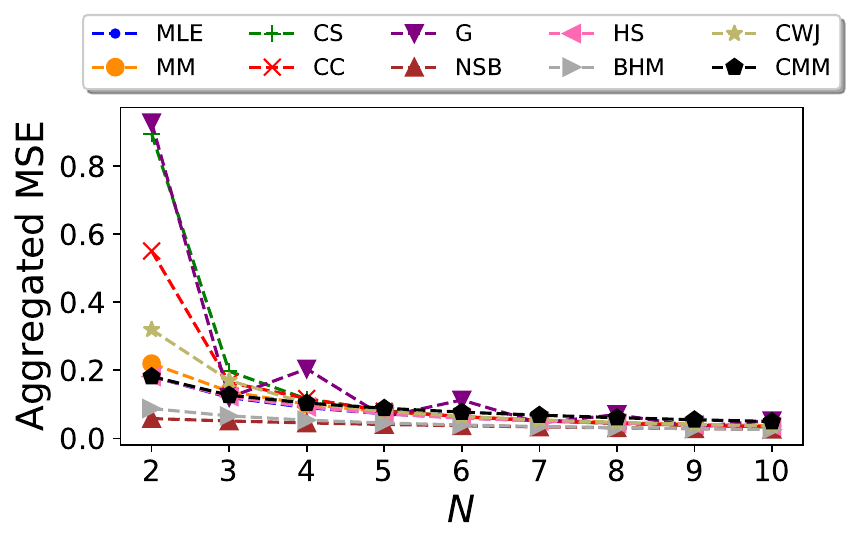}
\caption{Aggregated mean squared error of the entropy estimators for Markovian binary sequences as a function of the sequence size $N$.}
\label{fig:mse_total}
\end{figure}

\subsection{Undersampled Regime: Block Entropy}

Consider a sequence $S = X_1,\ldots,X_N$, where each $X_i = 0,1$ is a binary variable, with probabilities $P(X_i=1)= p$, $P(X_i=0)=1-p$. We group the sequence in blocks of size $n$, such that the $j$th-block is $B_j = (X_j,\ldots,X_{j+n-1})$. We denote by $\lbrace b_i \rbrace_{i=1,\dots,2^n}$ the set of all possible blocks. The total number of (overlapping) blocks that can be constructed out of a series of $N$ elements is $N_n = N-n+1$, whereas the total number of possible blocks is $L=2^n$. Hence, depending on the values of $n$ and $N$, the sequence formed by the $N_n$ blocks, $S_n = B_1,\ldots,B_{N_n}$, will be in an undersampled regime whenever $N_n\ll 2^n$.

The block entropy $H_n$ is defined by 
\begin{equation}
H_n = -\sum_{i=1}^{2^n} p(b_i)\ln(p(b_i)),
\end{equation}
where $p(b_i)$ is the probability of observing the block $b_i$. The important thing to notice here is that, even if the different outcomes $X_1,\dots, X_N$ of the binary variable $X$ are independent, the block sequence $B_1,\dots,B_{N_n}$ obeys a Markov process for $n\ge 2$.

 This Markovian property can be easily established by noticing that the block $B_j=(X_j,\ldots,X_{j+n-1})$ can only be followed by the block $B_{j+1}=(X_{j+1},\ldots,X_{j+n-1},1)$ with probability $p$ or by the block $B_{j+1}=(X_{j+1},\ldots,X_{j+n-1},0)$ with probability $1-p$. Therefore, the probability of $B_{j+1}$ depends only on the value of block $B_{j}$. In Appendix~\ref{sec:appB} we show that the dynamics of block sequences in the case that $X_i$ are i.i.d.~is equivalent to that of a new stochastic variable $Z$ that can take any of $L=2^n$ possible outcomes, $z_i=0,1,\ldots,2^n-1$, with the following transition probabilities for each state $z$:{
\begin{equation}
p(z_{k}|z_i) = \begin{cases}
 1-p, &\quad \text{if } z_{k} = 2z_i \,(\text{mod }2^n), \\
 p, &\quad \text{if } z_{k} = 2z_i \,(\text{mod }2^n) + 1,\\
 0, &\quad \text{otherwise}.\\
\end{cases}
\label{eq:trans_probs}
\end{equation}}
These types of Markovian systems have been related to Linguistics and Zipf's law~\citep{PhysRevLett.74.4559}.

The previous result can be generalized. If the original sequence $X_1,\dots,X_N$ is Markovian of order $m \geq 1$, then the dynamics of the block sequences $B_1,\dots,B_{N_n}$ are also Markovian of order $1$, for $n \geq m$.

It is well known~\cite{cover} that the block entropy, when the original sequence $S$ is constructed out of i.i.d.~binary variables, obeys
\begin{equation}
H_n = nH_1,
\label{eq:block_entropy}
\end{equation}
where $H_1$ can be calculated using Equation~\eqref{eq:H_binary} with $p(1)=p$ and $p(0)=1-p$. Therefore, the entropy rate is constant.

We want to compare now the performance of the different estimators defined before when computing the block entropy. In this case, we cannot use an expression equivalent to Equation~\eqref{eq:hk}, summing over all sequences $S_n$, since the number of possible sequences is $(2^n)^{N_n}$, and it is not possible to enumerate all the sequences even for relatively small values of $n$ and $N_n$. As an example, we employ in our numerical study $N_n=20$ and $n=6$, for which the total number of possible sequences is $2^{120}$. Therefore, we use the sample mean $\mu_M[\hat{H}_n]$ and the sample variance $s_M^2[\hat{H}_n]$ as unbiased estimators to the expected value $\langle \hat{H}_n\rangle$ and the variance $\sigma^2[\hat{H}_n]$, respectively. After generating a sample of $M$ independent sequences $S_n^i$, $i=1,\dots,M$, and computing the estimator $\hat{H}_n(S_n^i)$ for each of the sequences, those statistics are computed as
\begin{equation}\label{eq:Hmean}
\begin{split}
\mu_M[\hat{H}_n] &= \dfrac{1}{M} \sum_{i=1}^{M}\hat{H}_n(S_n^i),\\
s_M^2[\hat{H}_n] &= \frac{1}{M-1}\sum_{i=1}^{M} (\hat{H}_n(S_n^i)-\mu_M[\hat{H}_n] )^2.
\end{split}
\end{equation}
Using Equations~(\ref{eq:block_entropy}) and (\ref{eq:Hmean}) we can calculate the bias $B_n=\mu_M[\hat{H}_n]-H_n$, the standard deviation $s_M[\hat{H}_n]$, and the mean squared error $s_M^2[\hat{H}_n]+B_n^2$. In the following, we set $M=10^4$ for our simulations.

In Figure~\ref{fig:N=20}, we show plots of $B_n$ and $s_M[\hat{H}_n]$ as a function of $p$ ranging from $0.02$ to $0.5$ with step $\Delta p = 0.02$, for $N_n=20$. We find that the CC estimator performs remarkably well in terms of bias and we highlight its robustness. Unlike the other estimators, which display significant variations in their bias as $p$ changes, the CC estimator remains approximately constant at a low value. However, the CC estimator presents a high standard deviation, whereas the MLE and MM exhibit the lowest standard deviation.
For the majority of estimators considered, we observe that the ones with higher bias are the ones with lower deviation. An exception is the HS estimator.
\begin{figure}
\subfloat{\includegraphics[width=1\columnwidth]{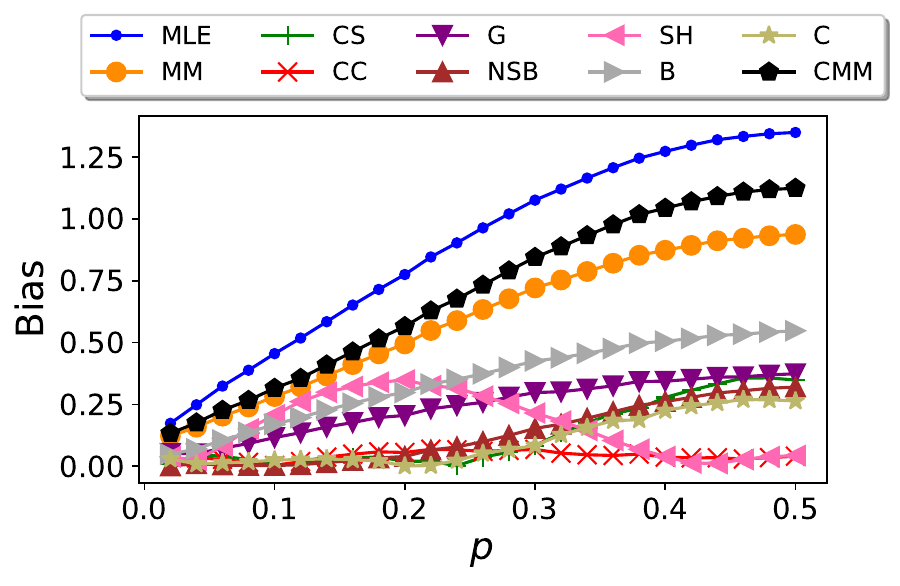}}
 \label{subfig:bias_N=20}\\
\subfloat{\includegraphics[width=1\columnwidth]{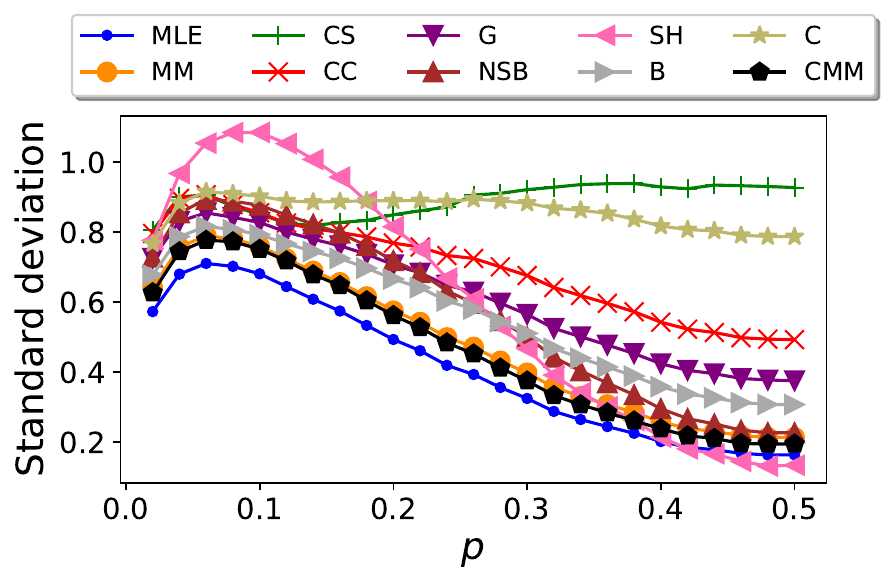}}
 \label{subfig:sigma_N=20}
\hfill
 \caption{Bias (top) and standard deviation (bottom) of the entropy estimators, when applied to Markovian sequences of length $N=20$ and $L=2^6$, generated from the transition probabilities given by Equation~\eqref{eq:trans_probs}, as functions of $p$, which vary from $0.02$ to $0.5$ with step $\Delta p = 0.02$. By construction, the plot is symmetric around $p=0.5$.} 
 \label{fig:N=20}
\end{figure}

To analyze the changes in the overall performances of the estimators with different values of $N$, we calculated the aggregated bias as
\begin{equation}
\overline{B}_n =\Delta p \sum_p |B_n(p)|.
\end{equation}
Similarly, we calculated the aggregated standard deviation as
\begin{equation}
\overline{s}_n =\Delta p \sum_p s_M[\hat{H}_n](p),
\end{equation}
and the aggregated mean squared error as
\begin{equation}
\overline{\text{MSE}}_n =\Delta p \sum_p (s_M^2[\hat{H}_n](p)+B_n(p)^2).
\end{equation}
The resulting plots are shown in Figures~\ref{fig:bias_block}--\ref{fig:mse_block}, respectively.
\begin{figure}[H]
\includegraphics[width=1\columnwidth]{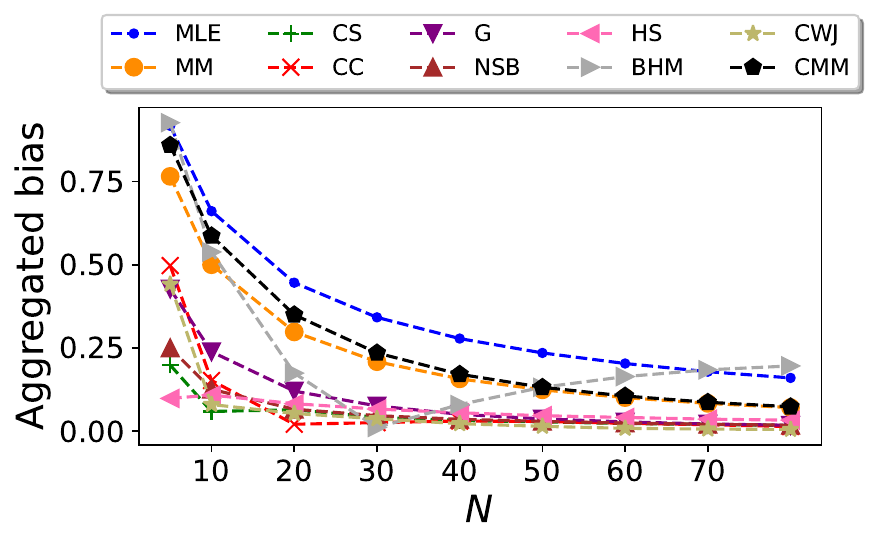}
\caption{Aggregated bias of the entropy estimators for Markovian sequences in the undersampled regime with $L=2^6$, generated from the transition probabilities given by Equation~\eqref{eq:trans_probs}, as a function of the sequence size $N$.}
\label{fig:bias_block}
\end{figure}
\unskip

\begin{figure}[H]
\includegraphics[width=1\columnwidth]{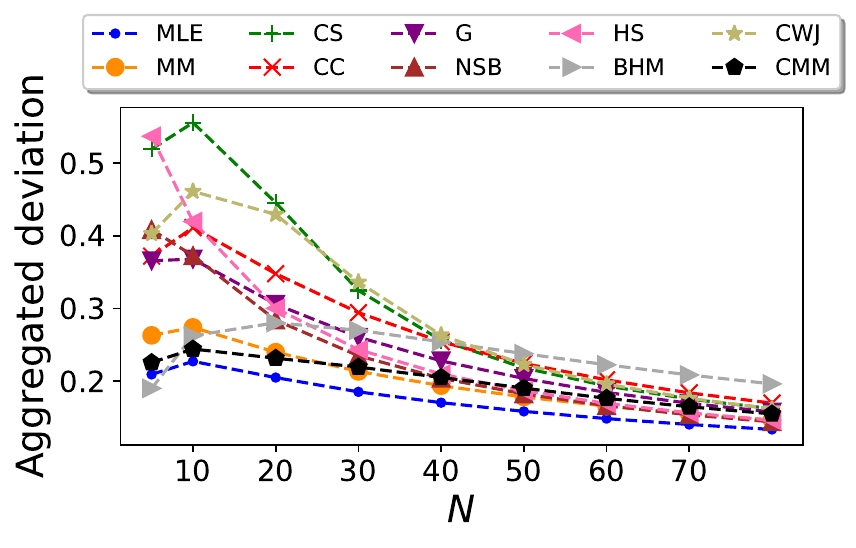}
\caption{Aggregated standard deviation of the entropy estimators for Markovian sequences in the undersampled regime with $L=2^6$, generated from the transition probabilities given by Equation~\eqref{eq:trans_probs}, as a function of the sequence size $N$.}
\label{fig:sigma_block}
\end{figure}
\unskip

\begin{figure}[H]
\includegraphics[width=1\columnwidth]{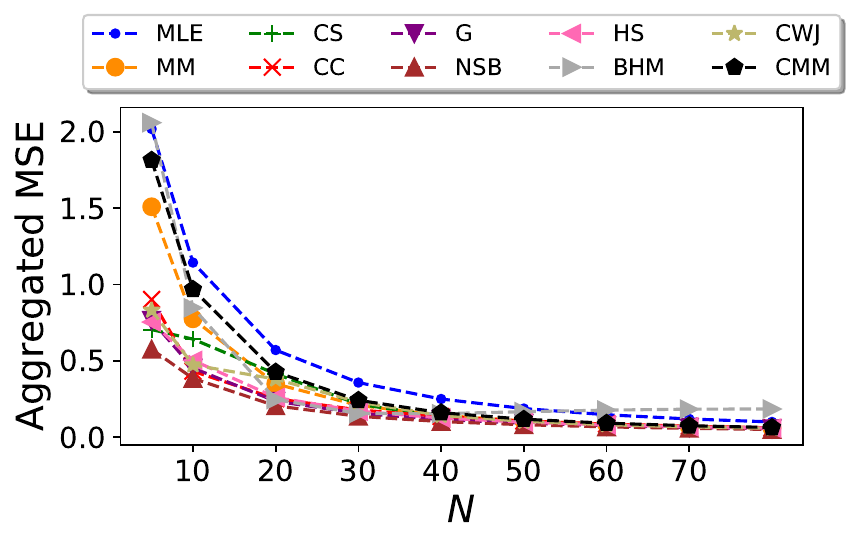}
\caption{Aggregated mean squared error of the entropy estimators for Markovian sequences in the undersampled regime with $L=2^6$, generated from the transition probabilities given by Equation~\eqref{eq:trans_probs}, as a function of the sequence size $N$.}
\label{fig:mse_block}
\end{figure}

It was expected that the total bias of the estimators would decrease by increasing $N$, and in Figure~\ref{fig:bias_block} it can be seen that this is indeed the case for all estimators except for the BHM estimator. Surprisingly, the bias of this estimator follows a typical pattern of decreasing as the sample size increases, just like the other estimators. However, it takes an unexpected turn starting at $N=20$, as it begins to increase once more.
A possible reason for this behaviour is that the BHM estimator is designed to minimize the MSE.

Similarly to the results obtained for the binary Markovian case, the CC estimator demonstrates in Figure~\ref{fig:bias_block} excellent performance when solely evaluating bias. Even though its performance for a data size of $N=5$ is not outstanding, it begins to outperform all but the CS, CWJ, and HS estimators starting at $N=10$, and from that point onward, the CC estimator consistently ranks among the top-performing estimators, together with the NSB and CWJ estimators.

By comparing Figures~\ref{fig:bias_block} and \ref{fig:sigma_block}, it can be seen that there is a certain balance: an estimator with a higher bias usually has a lower deviation when compared to others.
This is clearly the case for the MLE and MM estimators, as they are the two with the worst performances in terms of bias, but they have the lowest aggregated standard deviation for most of the data sizes considered.

In this interplay between bias and standard deviation observed for most of the entropy estimators considered here, the NSB estimator is the one that presents the best performance when considering both statistics.
From Figure~\ref{fig:mse_block}, it is clear that this estimator shows the lowest aggregated mean squared error, although just from $N=20$ the difference with other estimators like the CC or the G becomes vanishingly small.

{It can be seen in Figures~\ref{fig:bias_block}--\ref{fig:mse_block} that the performance of the CMM estimator is very similar to MM's performance, especially for large values of $N$. This suggests that for Markovian systems defined by the transition probabilities given by Equation~\eqref{eq:trans_probs}, the correction introduced in Equation~\eqref{eq:new} is not significant, particularly in the limit of large $N$.}

%%%%%%%%%%%%%%%%%%%%%%%%%%%%%%%%%%%%%%%%%%
\section{\label{sec:conclusions} Discussion}

We have made a detailed comparison of nine of the most widely used entropy estimators when applied to Markovian sequences. {We have also included in this analysis a new proposed estimator, motivated by the results presented in ref.~\cite{econometrics7020017}.}
One crucial difference in the way these estimators are constructed is that only the correlation coverage-adjusted estimator \cite{juan} {and the corrected Miller--Madow estimator take} into account the order in which the elements appear in the sequence. 
To calculate {the CC} estimator, it is necessary to know the entire history of the sequence {, and the computation of the CMM estimator requires the calculation of the transition probabilities. On the contrary,} for all other estimators, it is sufficient to know the number of times that each element is present in the sequence, independently of the position in which they appear. Remarkably, this novel {approach to} 
the issue of entropy estimation} allows us to reduce the bias, even in undersampled regimes.
Unfortunately, {both of these estimators present} large dispersion, which reduces their overall~quality.

We have found that, when dealing with Markovian sequences, on average, the Nemenman--Shafee--Bialek estimator \cite{nsb,nsb2,nsb3} outperforms the rest when taking into account both the bias and the standard deviation for both analyzed cases, namely, binary sequences and an undersampled regime. Ref.~\cite{contreras} presented a similar analysis but for uniformly distributed sequences of bytes and bites, and concluded that the estimator with the lowest mean squared error was the Shrinkage estimator \cite{shr}. 
Hence, when choosing a reliable estimator, it is not only important to consider the amount of data available, but also whether correlations might be present in the sequence.

Further analyses should consider Markovian sequences of higher order \cite{raftery1985model,strelioff2007inferring}. Another interesting topic would be systems described with
continuous variables \cite{bercher2000estimating,feutrill2021review}, where the presence of noise is particularly
important. Finally, we stress that there are alternative entropies
not considered here \cite{beck}, for which the existence of accurate estimators
is still an open question. Finally, an exciting possibility would be a comparative study of estimators valid for more than one random variable or probability distributions, leading, respectively, to mutual information~\cite{kraskov2004estimating,walters2009estimation} and relative entropy \cite{piga2023bayesian,minculete2023some,camaglia2023bayesian}.

%%%%%%%%%%%%%%%%%%%%%%%%%%%%%%%%%%%%%%%%%%
%\section{Conclusions}

%This section is not mandatory, but can be added to the manuscript if the discussion is unusually long or complex.

%%%%%%%%%%%%%%%%%%%%%%%%%%%%%%%%%%%%%%%%%%
%\section{Patents}

%This section is not mandatory, but may be added if there are patents resulting from the work reported in this manuscript.

%%%%%%%%%%%%%%%%%%%%%%%%%%%%%%%%%%%%%%%%%%
\vspace{6pt} 

%%%%%%%%%%%%%%%%%%%%%%%%%%%%%%%%%%%%%%%%%%
%% optional
%\supplementary{The following supporting information can be downloaded at: \linksupplementary{s1}, Figure S1: title; Table S1: title; Video S1: title.}

%%%%%%%%%%%%%%%%%%%%%%%%%%%%%%%%%%%%%%%%%%
\section{Acknowledgments}
Partial financial support has been received from the Agencia Estatal de Investigaci\'on (AEI, MCI, Spain) MCIN/AEI/10.13039/501100011033 and Fondo Europeo de Desarrollo Regional (FEDER, UE) under Project APASOS (PID2021-122256NB-C21) and the María de Maeztu Program for units of Excellence in R\&D, grant CEX2021-001164-M.

%\newpage
%\clearpage

%%%%%%%%%%%%%%%%%%%%%%%%%%%%%%%%%%%%%%%%%%

\onecolumngrid
\appendix
\section[\appendixname~\thesection]{}
%\subsection[\appendixname~\thesubsection]{}
\label{sec:appA}
In {Appendix} 
~\ref{sec:appA},  we introduce a new estimator $\hat{h}_0$ of $-p\ln(p)$ based on the number of observations made prior to the occurrence of the result $x$ with probability $p$. We improve this estimator by including all contributions resulting from the shuffling of the original series. Additionally, we show that this improved estimator $\hat{h}$ has been used as a starting point to construct different estimators proposed in the literature.

Let $x$ be a possible value, with probability $p$, of a random variable $X$. We make independent repetitions of $X$ and define a new random variable $K$ as the number of repetitions until the result $x$ occurs for the first time. This random variable follows a geometric distribution: $P(K=k) = p(1-p)^{k-1},\,k\ge1$.
Let us consider the following random variable
\begin{equation}
R = \begin{cases}
0 & \text{if } K = 1,\\
 \dfrac{1}{K-1} & \text{if } K \geq 2.
\end{cases}
\end{equation}
The average value of $R$ is
\begin{equation}
\langle R\rangle = \sum_{k=2}^{\infty} \dfrac{p(1-p)^{k-1}}{k-1} = -p\ln(p),
\label{eq:mean_r}
\end{equation}
where we have used a known series expansion of the logarithm function. Hence, $R$ is an unbiased estimator of $-p\ln(p)$ {\cite{montgomerysmith2014unbiased}}. By adding similar random variables $R_i$ for each possible result $x_i,\,i=1,\dots,L$, we can obtain a random variable whose average value is Shannon's entropy. This is not a contradiction with the statement that there is no known unbiased estimator of the entropy for a series of finite length, as a proper evaluation of this estimator requires the possibility of repeating infinite times the random variable. If the maximum allowed number of repetitions is $N$, we must modify the definition of the random variable as
\begin{equation}
R_{_N} = \begin{cases}
0 & \text{if } K = 1, \text{ or }K>N\\
 \dfrac{1}{K-1} & \text{if } 2 \leq K \leq N.
\end{cases}
\end{equation}
It turns out that $R_{_N}$ is negatively biased because
\begin{equation}
\langle R_{_N}\rangle=\sum_{k=2}^{N} \dfrac{p(1-p)^{k-1}}{k-1}= -p \log(p)-p(1-p)^N\Phi(1-p,1,N),
\label{eq:mean_r2}
\end{equation}
where $\Phi(z,1,N)=\sum_{k=0}^{\infty}\dfrac{z^k}{N+k}$ is Lerch's transcendent function.

Based on this result, we introduce the following estimator $\hat{h}_0$ for $-p\log(p)$: given a series $S=X_1,\dots,X_N$ in which the symbol $x$ appears $n$ times, we count the set of distances $(k_1,k_2,\dots,k_n)$ between successive appearances of the symbol $x$ and then define:
\begin{equation}
\hat{h}_0(S)=\frac{1}{n}\sum_{j=1}^n\frac{\Theta(k_j-1)}{k_j-1}.
\end{equation}
The $\Theta$ function implements the condition $k_j\ge 2$ and the condition $k_j\le N$ appears naturally because of the number of data in the series. As the different points in the series are the results of independent repetitions of the random variable $X$, it is possible to reshuffle all points and still obtain a representative series of the process, whereas the usual MLE estimator is insensitive to this reshuffling, as it only depends on the number of appearances $n$, the estimator $\hat{h}_0(S)$ does depend on the order of the sequence. Therefore, it is possible to improve the statistics of this estimator by including all contributions of the $N!$ possible permutations of the $N$ terms of the original series. If $(k_1^{(i)},\dots, k_n^{(i)})$ is the set of distances between successive appearances of the $x$ symbol in the $i$-th permutation, then we define the improved estimator
\begin{equation}\label{eq:hathS1}
\hat{h}(S)=\frac{1}{N!}\sum_{i}\frac{1}{n}\sum_{j=1}^n\frac{\Theta(k_j^{(i)}-1)}{k_j^{(i)}-1},
\end{equation}
where the sum over $i$ runs over all possible permutations of the original sequence. Our main result is to prove that this estimator can be written in terms only of $n$ and $N$, namely
\begin{equation}\label{mainresult}
\hat{h}(S)=\frac{n}{N}\sum_{k=n+1}^N\frac{1}{k-1}=\dfrac{n}{N}(\psi(N)-\psi(n)),
\end{equation}
where $\psi(z)$ is the digamma function, the logarithmic derivative of the gamma function. See proof in Appendix~\ref{subsec:1} 
 of Appendix~\ref{sec:appA}.

The average value of $\hat{h}(S)$ is given by
\begin{equation}
\langle \hat{h}\rangle = \sum_{n=0}^NP(n) \dfrac{n}{N}(\psi(N)-\psi(n))
\end{equation}
where
\begin{equation}\label{pbinom}
P(n) = \binom{N}{n} p^n (1-p)^{N-n},
\end{equation}
is the probability that the element $x$ appears $n$ times in a sequence of length $N$. As proven in Appendix~\ref{subsec:2}, the average value is
\begin{equation}
\langle \hat{h}\rangle =-p \log(p)-p(1-p)^N\Phi(1-p,1,N),
\end{equation}
which proves that $\hat{h}$ is an unbiased estimator of $\langle R_{_N}\rangle$. 

Repeating this same procedure for every $x_i$ in the sequence with $n_i > 0$, we arrive at the entropy estimator
\begin{equation}\label{eq:Hr}
\HR =\sum_{i=1}^L \dfrac{n_i}{N} (\psi(N)-\psi(n_i)),
\end{equation}
whose bias is the sum of the biases associated with each value of the random variable
\begin{equation}
B[\HR] = -\sum_{i=1}^L p(x_i)(1-p(x_i))^N\Phi(1 -p(x_i),1,N).
\label{eq:biasHr}
\end{equation} 
As proven in the supplementary material of \cite{chao}, $\hat{h}(S)$ has been used as a starting point to construct the estimators CWJ {amongst others \cite{chao,vinck,zhang,Sch_rmann_2004}.}
For example, ref.~\cite{zhang} proposes to correct $\HR$ in Equation~\eqref{eq:Hr} by subtracting to this definition the bias in Equation~\eqref{eq:biasHr}, replacing the values of the unknown probabilities by their estimated frequencies, {\linebreak}$p(x_i)\to \dfrac{n_i}{N}$. In~\cite{vinck}, the correcting bias subtraction is estimated using a Bayesian approach. Finally, in~\cite{chao}, the authors recognized that the greatest contribution to the bias must come from the outcomes that do not appear in the sequence. Hence, they propose to correct $\HR$ by using an improved Good--Turing formula~\cite{good} to account for the missing elements in the sequence, leading to the estimator given by Equation~\eqref{eq:chao}.

The novel strategy presented here to introduce the estimator $\HR$ emphasizes its relation with the geometric distribution and provides further insight into its significance.

\subsection{Proof of Equation~(\ref{mainresult})}
\label{subsec:1}

We prove it in three steps:\\
\textbf{Step 1:}
Note that not all permutations give a different set $(k_1^{(i)},\dots, k_n^{(i)})$. There are, in fact, only $N\choose n$ permutations that differ in the value of the sequence $(k_1^{(i)},\dots, k_n^{(i)})$, corresponding to the selection of the $n$ locations of the $x$ symbol in the sequence. Therefore, we can simplify the expression for the estimator as
\begin{equation}\label{eq:hathS2}
\hat{h}(S)=\frac{1}{n}\frac{1}{{N\choose n}}\sum_{i=1}^{{N\choose n}}\sum_{j=1}^n\frac{\Theta(k_j^{(i)}-1)}{k_j^{(i)}-1},
\end{equation}
where the sum over $i$ now runs over the permutations that give rise to a different set of numbers $(k_1^{(i)},\dots, k_n^{(i)})$.

\noindent\textbf{{Step~2:}} We show that the double sum in Equation~\eqref{eq:hathS2} can be written as a function of $n$ and $N$ only,
\begin{equation}\label{eq:prop1}
\sum_{i=1}^{{N\choose n}}\sum_{j=1}^n\frac{\Theta(k_j^{(i)}-1)}{k_j^{(i)}-1}\equiv R(n,N).
\end{equation}
where
\begin{equation}\label{eq:prop2}
R(n,N)= n\sum_{k=2}^{N-n+1}\binom{N-k}{n-1}\dfrac{1}{k-1}.
\end{equation}
We prove this relation by mathematical induction.
Consider the case $n=1$. The $N$ permutations that differ in the value of $k$ correspond to the appearance of the symbol $x$ in the first term of the series ($k=1$), the second term of the series ($k=2$), and so on up to the $N$-th term ($k=N$). The sum in the left-hand-side of Equation~\eqref{eq:prop1} is
\begin{equation}
\sum_{k=2}^{N} \dfrac{1}{k-1},
\label{eq:R1}
\end{equation}
which coincides with $R(1,N)$, defined in Equation~\eqref{eq:prop2}.

Assume now that Equation~\eqref{eq:prop2} is valid up to $1 \leq n \leq N-1$, and let us evaluate $R(n+1,N)$. Consider all possible permutations in a sequence of length $N$ that start with $(x,\ldots)$. The total contribution of these sequences to the value of $R(n+1,N)$ is the same as having all permutations of a sequence of length $N-1$ with $n$ occurrences of $x$ (notice that the contribution of the first appearance of $x$ is equal to $0$). 

We then consider all $\binom{N-2}{n}$ permutations that start with $(0,x,\ldots)$, where with "$0$" we indicate any value which is not equal to $x$. That first appearance of $x$ will contribute with a term equal to $1$ for each of the permutations, and the rest will contribute the same as having all permutations of a sequence of length $N-2$ with $n$ occurrences of $x$. Following this procedure, we have that

%\centering %% If there is a figure in wide page, please release command \centering
\begin{equation}
R(n+1,N) = R(n,N-1) + \binom{N-2}{n}\dfrac{1}{2-1} + R(n,N-2) + \ldots + \dfrac{1}{(N-n)-1}+R(n,n),
\label{eq:Rn+1}
\end{equation}
where the last two terms correspond to the contribution of the permutation that has all $n+1$ occurrences of $x$ at the end.

Given that we are assuming that \begin{equation}
R(n+1,N) = \sum_{k=2}^{N-n-1}\binom{N-k}{n}\dfrac{1}{k-1} + \dfrac{1}{N-n-1} + n\sum_{k=2}^{N-n}\binom{N-k-1}{n-1}\dfrac{1}{k-1} + n\sum_{j=2}^{N-n-1}\sum_{k=2}^{N-j-n+1}\binom{N-j-k}{n-1}\dfrac{1}{k-1}.
\label{eq:Rn+1_2}
\end{equation}
%\end{adjustwidth}
Changing the order of summation of the last term in Equation~\eqref{eq:Rn+1_2}, we can write it as
%\begin{adjustwidth}{-\extralength}{0cm}
\begin{align}
\begin{split}
\sum_{k=2}^{N-n-1}\dfrac{1}{k-1}\sum_{j=2}^{N-k-n+1}\binom{N-j-k}{n-1} &= \sum_{k=2}^{N-n-1}\dfrac{1}{k-1}\sum_{u=0}^{N-k-n-1}\binom{u+n-1}{n-1} = \sum_{k=2}^{N-n-1}\dfrac{1}{k-1}\binom{N-k-1}{n},
\end{split}
\end{align}
%\end{adjustwidth}
where the last equality is due to Fermat's combinatorial identity (mostly known as the hockey-stick identity). Hence, Equation~\eqref{eq:Rn+1_2} can be written as
\begin{align}
\begin{split}
R(n+1,N) &= \sum_{k=2}^{N-n-1}\binom{N-k}{n}\dfrac{1}{k-1}+\dfrac{1}{N-n-1}+n\sum_{k=2}^{N-n}\binom{N-k-1}{n-1}\dfrac{1}{k-1}+n\sum_{k=2}^{N-n-1}\binom{N-k-1}{n}\dfrac{1}{k-1} \\
&= \sum_{k=2}^{N-n-1}\binom{N-k}{n}\dfrac{1}{k-1} + (n+1)\dfrac{1}{N-n-1}+n\sum_{k=2}^{N-n-1} \left( \binom{N-k-1}{n}+\binom{N-k-1}{n-1} \right) \dfrac{1}{k-1}.
\end{split}
\end{align}
Using Pascal's identity
\begin{equation}
\binom{N-k-1}{n}+\binom{N-k-1}{n-1} = \binom{N-k}{n}.
\end{equation}
 we obtain,
\begin{align}
\begin{split}
R(n+1,N) &= (n+1)\dfrac{1}{N-n-1}+(n+1)\sum_{k=2}^{N-n-1}\binom{N-k}{n}\dfrac{1}{k-1} \\
&= (n+1)\sum_{k=2}^{N-n}\binom{N-k}{n}\dfrac{1}{k-1},
\end{split}
\end{align}
which proves Equation~\eqref{eq:prop2} for $1\leq n \leq N$.

\textbf{{Step~3:}} We show that $\hat{h}$ can finally be written as
\begin{equation}
\hat{h} = \dfrac{1}{n}\dfrac{1}{\binom{N}{n}} R(n,N) = \frac{n}{N}\sum_{k=n+1}^N\frac{1}{k-1}=\dfrac{n}{N}(\psi(N)-\psi(n)),
\label{eq:Rest2}
\end{equation}
where $\psi$ is the digamma function.

The proof again uses mathematical induction. Consider first the case $n=1$. From Equations~(\ref{eq:hathS1})--(\ref{eq:prop1}), we derive
\begin{equation}
\dfrac{1}{\binom{N}{1}} R(1,N) = \dfrac{1}{N}\sum_{k=2}^{N}\dfrac{1}{k-1} = \dfrac{1}{N}(\psi(N)-\psi(1)),
\end{equation}
where the last equality is a known identity of the Harmonic numbers.

Consider now that Equation~\eqref{eq:Rest2} holds for $1\leq n \leq N-1$. Let us evaluate the case $n+1$:
%\begin{adjustwidth}{-\extralength}{0cm}
\begin{align}
\begin{split}
\dfrac{1}{n+1}\dfrac{1}{\binom{N}{n+1}}R(n+1,N) &= \dfrac{1}{\binom{N}{n+1}}\sum_{k=2}^{N-n}\binom{N-k}{n}\dfrac{1}{k-1} = \dfrac{n+1}{N}\dfrac{1}{\binom{N-1}{n}}\sum_{k=2}^{N-n}\dfrac{N-k}{n}\binom{N-k-1}{n-1}\dfrac{1}{k-1} \\
&= \dfrac{n+1}{N} \left( \dfrac{1}{\binom{N-1}{n}} \dfrac{N-1}{n} \sum_{k=2}^{N-n}\binom{N-1-k}{n-1}\dfrac{1}{k-1}-\dfrac{1}{\binom{N-1}{n}} \dfrac{1}{n}\sum_{k=2}^{N-n}\binom{N-k-1}{n-1} \right).
\end{split}
\end{align}
%\end{adjustwidth}
Notice that, given our induction hypothesis,
\begin{equation}
\dfrac{1}{\binom{N-1}{n}}\sum_{k=2}^{N-n}\binom{N-1-k}{n-1}\dfrac{1}{k-1} = \dfrac{1}{n}\dfrac{1}{\binom{N-1}{n}}R(n,N-1) =\dfrac{n}{N-1}(\psi(N-1)-\psi(n)).
\end{equation}
Hence,
\vspace{-9pt}
\begin{align}
\begin{split}
\dfrac{1}{n+1}\dfrac{1}{\binom{N}{n+1}}R(n+1,N) &= \dfrac{n+1}{N} \left(\psi(N-1)-\psi(n) - \dfrac{1}{\binom{N-1}{n}} \dfrac{1}{n} \sum_{j=0}^{N-n-2}\binom{j+n-1}{n-1} \right) \\
&= \dfrac{n+1}{N} \left(\psi(N-1)-\psi(n) - \dfrac{1}{\binom{N-1}{n}} \dfrac{1}{n} \binom{N-2}{n} \right) \\
&= \dfrac{n+1}{N} \left(\psi(N-1)-\psi(n) - \dfrac{1}{n} + \dfrac{1}{N-1} \right) \\
&=\dfrac{n+1}{N} \left(\psi(N)-\psi(n+1) \right),
\end{split}
\end{align}
where for the last equality we have used the known property of the digamma function: $\psi(z+1)=\psi(z)+1/z$.

\subsection{Calculation of the Average $\langle \hat{h}(S)\rangle$}
\label{subsec:2}
The average value of the estimator $\hat{h}$ is
\begin{equation}
\langle \hat{h}\rangle =\sum_{n=0}^{N}P(n)\frac{1}{n}\frac{1}{{N\choose n}}R(n,N)
\end{equation}
where $P(n)$ is given in Equation~\eqref{pbinom} and we will use the expression given in Equation~\eqref{eq:prop2} for $R(n,N)$. Hence,
\begin{align}
\begin{split}
\langle \hat{h}\rangle&= \sum_{n=1}^{N-1} \binom{N}{n} p^n (1-p)^{N-n}\sum_{k=2}^{N-n+1} \dfrac{1}{\binom{N}{n}}\binom{N-k}{n-1} \dfrac{1}{k-1} \\
&= \sum_{n=0}^{N-2}\sum_{k=2}^{N-n} \dfrac{1}{k-1} p^{n+1}(1-p)^{N-n-1}\binom{N-k}{n},
\end{split}
\end{align}
changing the order of summation we have,
\begin{align}
\begin{split}
\langle \hat{h}\rangle&= \sum_{k=2}^{N}\sum_{n=0}^{N-k} \dfrac{1}{k-1} p^{n+1}(1-p)^{N-n-1}\binom{N-k}{n} \\
&= \sum_{k=2}^{N}\dfrac{1}{k-1}p(1-p)^{k-1}\sum_{n=0}^{N-k} \binom{N-k}{n} p^n(1-p)^{N-k-n},
\end{split}
\end{align}
the second sum of the equation above is just the binomial expansion of $(p+1-p)^{N-k}$ which is equal to $1$. Then,
\begin{equation}
\langle \hat{h}\rangle = \sum_{k=2}^{N}\dfrac{1}{k-1}p(1-p)^{k-1} = -p\log(p)-p(1-p)^N\Phi(1-p,1,N) .
\end{equation}

\section[\appendixname~\thesection]{}
\label{sec:appB}
Consider a Markovian sequence with $L=2^n$ possible outcomes, $z_i=0,1,\ldots,2^n-1$, defined by the following transition probabilities:
{
\begin{equation}
p(z_{k}|z_i) = \begin{cases}
 1-p, &\quad \text{if } z_{k} = 2z_i \quad(\text{mod }2^n), \\
 p, &\quad \text{if } z_{k} = 2z_i + 1\quad(\text{mod }2^n) ,\\
 0, &\quad \text{otherwise}.\\
\end{cases}
\label{eq:trans_probs2}
\end{equation}}
We can write any $z_i$ in base $2$ as
\begin{equation}
z_i = X_{1} 2^{n-1} + X_{2} 2^{n-2}+ \ldots + X_{n},
\end{equation}
where each $ X_{j}$ is either $0$ or $1$. Then, we can represent the state $z_i$ as a binary string of size $n$: $z_i\equiv ( X_{1},\ldots, X_{n})$.
Hence,
\begin{equation}
2z_i = X_{1} 2^n + X_{2} 2^{n-1} + \ldots + X_{n} 2.
\label{eq:2x}
\end{equation}
Reducing the modulo $2^n$ Equation~\eqref{eq:2x}, we have
\begin{equation}
2z_i\,(\text{mod } 2^n) = X_{2} 2^{n-1}+\ldots+ X_{n} 2 + 0\equiv( X_{2},\ldots, X_{n},0)
\label{eq:m0}
\end{equation}
and
\begin{equation}
2z_i\,(\text{mod } 2^n)+1 = X_{2} 2^{n-1}+\ldots+ X_{n} 2 + 1\equiv( X_2,\ldots, X_n,1)
\label{eq:m1}
\end{equation}
Hence, the dynamics of this system are equivalent to a block sequence in which the block $( X_1,\ldots, X_{n})$ can only be followed by the block $( X_2,\ldots, X_n,0)$ with probability $1-p$ or by $( X_2,\ldots, X_n,1)$ with probability $p$, coincident with Equation~\eqref{eq:trans_probs2}.

\end{document}